\documentclass[aps]{revtex4}
\usepackage{graphicx}

\begin{document}
\newcommand{\be}{\begin{equation}}
\newcommand{\ee}{\end{equation}}
\newcommand{\bea}{\begin{eqnarray}}
\newcommand{\eea}{\end{eqnarray}}
\newcommand{\ba}{\begin{array}}
\newcommand{\ea}{\end{array}}
\newcommand{\Dslash}{D\hspace*{-0.23cm}/\,}

\title{Quark Matter}

\author{Thomas Sch\"afer}
\affiliation{  Department of Physics,
     North Carolina State University,
     Raleigh, NC 27695 \\
     and Riken-BNL Research Center, 
     Brookhaven National Laboratory, Upton, NY 11973 }

\begin{abstract}
In these lectures we provide an introduction to the 
theory of QCD at very high baryon density. We begin 
with a review of some aspects of quantum many-body 
system that are relevant in the QCD context. We also
provide a brief review of QCD and its symmetries. The 
main part of these lectures is devoted to the phenomenon 
of color superconductivity. We discuss the use of weak 
coupling methods and study the phase structure as a 
function of the number of flavors and their masses. 
We also introduce effective theories that describe 
low energy excitations at high baryon density. Finally, 
we use effective field theory methods in order to study 
the effects of a non-zero strange quark mass.
\end{abstract}

\maketitle
\newpage
\tableofcontents
\newpage
\section{Introduction}
\label{sec_intro}

 In these lectures we wish to provide an introduction 
to recent work on the phase structure of QCD at non-zero
baryon density. This work is part of a larger effort 
to understand the behavior of matter under ``extreme''
conditions such as very high temperature or very large 
baryon density. There are several motivations for 
studying extreme QCD:

\begin{itemize}

\item{Extreme conditions exist in the universe: About
$10^{-5}$ sec after the big bang the universe passed
through a state in which the temperature was comparable
to the QCD scale. Much later, matter condensed into stars.
Some of these stars, having exhausted their nuclear fuel,
collapse into compact objects called neutron stars. The
density at the center of a neutron star is not known
very precisely, but almost certainly greater or equal
to the density where quark degrees of freedom become
important.}

\item{Exploring the entire phase diagram is important
to understanding the phase that we happen to live in:
We cannot properly understand the structure of hadrons
and their interactions without understanding the underlying
QCD vacuum state. And we cannot understand the vacuum
state without understanding how it can be modified. }

\item{QCD simplifies in extreme environments: At scales
relevant to hadrons QCD is strongly coupled and we have
to rely on numerical simulations in order to test predictions
of QCD. In the case of large temperature or large baryon
density there is a large external scale in the problem.
Asymptotic freedom implies that the bulk of the system is
governed by weak coupling. As a result, we can study
QCD matter in a regime where quarks and gluons are indeed
the correct degrees of freedom.}

\item{Finally, extreme QCD tries to answer one of the 
simplest and most straightforward questions about the 
behavior of matter: What happens if we take a piece of 
material and heat it up to higher and higher temperature, 
or compress to larger and larger density?}

\end{itemize}

 There are several excellent text books and reviews articles 
that provide an introduction to QCD and hadronic matter at 
finite temperature \cite{Shuryak:1988,Kapusta:1989,LeBellac:1996}. 
In these lectures we will focus on matter at high baryon density 
and small or zero temperature. This is the regime of the 
``condensed matter physics'' of QCD \cite{Rajagopal:2000wf}. 
Ordinary condensed matter physics is concerned with the 
overwhelmingly varied appearance and rich phase diagram 
of matter composed of electrons and ions. All phases
of condensed matter ultimately derive their properties 
from the simple laws of quantum electrodynamics. We
expect, therefore, that the simple laws of QCD will 
lead to a phase diagram of comparable diversity. In fact,
since there is only one kind of electron, but several flavors 
and colors of quarks, we might expect new and unusual 
phases of matter never before encountered. 

 These lectures are organized as follows. In sections
\ref{sec_fl}-\ref{sec_sc} we review a number of simple 
many body systems that are relevant to the behavior of 
QCD matter in different regimes. In order to keep the 
presentation simple, and to make contact with well-known 
properties of other many body system, we phrase our 
discussion not in terms of quarks and gluons, but in 
terms of generic fermions and bosons interacting via 
short range range forces. In section \ref{sec_qcd} we 
provide a brief introduction to QCD and its symmetries. 
Sections \ref{sec_csc}-\ref{sec_ms} form the main 
part of these lectures. We introduce the phenomenon of
color superconductivity, study the phase structure in 
weak coupling, and introduce effective field theories that 
allow systematic calculations of the properties of dense 
QCD matter. Other aspects of high density QCD are 
discussed in the many excellent reviews on the subject
\cite{Rajagopal:2000wf,Alford:2001dt,Nardulli:2002ma,Reddy:2002ri}.

\section{Fermi liquids}
\label{sec_fl}
\subsection{Introduction}

  In this section we wish to study a system of non-relativistic
fermions interacting via a short-range interaction 
\cite{Abrikosov:1963,Hammer:2000xg}. The lagrangian is 
\be 
\label{l_4f}
{\cal L}_0 = \psi^\dagger \left( i\partial_0 +
 \frac{\nabla^2}{2m} \right) \psi 
 - \frac{C_0}{2} \left(\psi^\dagger \psi\right)^2 .
\ee
The coupling constant $C_0$ is related to the scattering 
length, $C_0=4\pi a/m$. Note that $C_0>0$ corresponds to 
a repulsive interaction, and $C_0<0$ is an attractive interaction.
The lagrangian equ.~(\ref{l_4f}) is invariant under the $U(1)$ 
transformation $\psi\to e^{i\phi}\psi$. The $U(1)$ symmetry 
implies that the fermion number 
\be
 N= \int d^3x\,\psi^\dagger \psi
\ee
is conserved. As a consequence, it is meaningful to study a 
system of fermions at finite density $\rho=N/V$. We will do 
this in the grand-canonical formalism. We introduce a chemical 
potential $\mu$ conjugate to the fermion number $N$ and 
study the partition function
\be 
\label{Z}
 Z(\mu,\beta) = {\rm Tr}\left[e^{-\beta(H-\mu N)}\right].
\ee
Here, $H$ is the Hamiltonian associated with ${\cal L}$ 
and $\beta=1/T$ is the inverse temperature. The trace in 
equ.~(\ref{Z}) runs over all possible states of the system, 
including all sectors of the theory with different particle
number $N$. The average number of particles for a given 
chemical potential $\mu$ and temperature $T$ is given 
by $\langle N\rangle =T(\partial \log Z)/(\partial \mu)$. 
At zero temperature the chemical potential is the energy
required to add one particle to the system. 

\begin{figure}
\includegraphics[width=16.0cm]{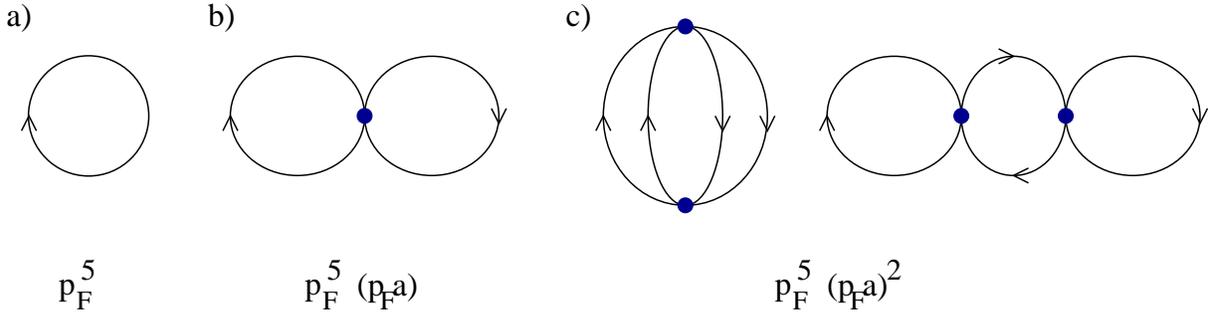}
\caption{\label{fig_fl}
Leading order Feynman diagrams for the ground state 
energy of a dilute gas of fermions interacting via
a short range potential.}
\end{figure}

 There is a formal resemblance between the partition function
equ.~(\ref{Z}) and the quantum mechanical time evolution 
operator $U=\exp(-iHt)$. In order to write the partition
function as a time evolution operator we have to identify 
$\beta\to it$ and add the term $-\mu N$ to the Hamiltonian. 
Using standard techniques we can write the time evolution
operators as a path integral \cite{Kapusta:1989,LeBellac:1996}
\be 
Z = \int D\psi D\psi^\dagger \exp\left(-\int_0^\beta d\tau 
  \int d^3x\, {\cal L}_E \right).
\ee
Here, ${\cal L}_E$ is the euclidean lagrangian
\be 
\label{l_4f_E}
{\cal L}_E = \psi^\dagger \left( \partial_\tau 
 - \mu - \frac{\nabla^2}{2m} \right) \psi 
 + \frac{C_0}{2} \left(\psi^\dagger \psi\right)^2 .
\ee
The fermion fields satisfy anti-periodic boundary 
conditions $\psi(\beta)=-\psi(0)$. Equation (\ref{l_4f_E}) is 
the starting point of the imaginary time formalism in 
thermal field theory. The chemical potential simply results 
in an extra term $-\mu\psi^\dagger\psi$ in the lagrangian. 
From equ.~(\ref{l_4f_E}) we can easily read off the free 
fermion propagator 
\be
S_{\alpha\beta}^0(p) = \frac{\delta_{\alpha\beta}}
  {ip_4+\mu-\frac{\vec{p}^{\, 2}}{2m}},
\ee
where $\alpha,\beta$ are spin labels. We observe that the 
chemical potential simply shifts the four-component of the 
momentum. This implies that we have to carefully analyze the
boundary conditions in the path integral in order to fix the 
pole prescription. The correct Minkowski space propagator is
\be
\label{s_ph} 
S^0_{\alpha\beta}(p) =
 \frac{\delta_{\alpha\beta}}
 {p_0-\epsilon_p+i\delta{\rm sgn}(\epsilon_p)}
 = \delta_{\alpha\beta}\left\{
 \frac{\Theta(p-p_F)}{p_0-\epsilon_p+i\delta}+
 \frac{\Theta(p_F-p)}{p_0-\epsilon_p-i\delta}
  \right\},\nonumber
\ee
where $\epsilon_p=E_p-\mu$, $E_p=\vec{p}^{\, 2}/(2m)$ and
$\delta\to 0^+$. The quantity $p_F=\sqrt{2m\mu}$ is called 
the Fermi momentum. We will refer to the surface defined 
by the condition $|\vec{p}|=p_F$ as the Fermi 
surface. The two terms in equ.~(\ref{s_ph}) have a simple 
physical interpretation. At finite density and zero temperature
all states with momenta below the Fermi momentum are 
occupied, while all states above the Fermi momentum are
empty. The possible excitation of the system are particles
above the Fermi surface or holes below the Fermi surface,
corresponding to the first and second term in 
equ.~(\ref{s_ph}). The particle density is given by
\be
\rho = \langle\psi^\dagger\psi\rangle =
 \int \frac{d^4p}{(2\pi)^4} S^0_{\alpha\alpha}(p)
 \left. e^{ip_0\eta}\right|_{\eta\to 0^+}
 = 2\int \frac{d^3p}{(2\pi)^3}\Theta(p_F-p)
 = \frac{p_F^3}{3\pi^2}.
\ee 
As a first simple application we can compute the energy 
density as a function of the fermion density. For free
fermions, we find 
\be
{\cal E} =  2\int \frac{d^3p}{(2\pi)^3}E_p\Theta(p_F-p)
 = \frac{3}{5}\rho\frac{p_F^2}{2m}.
\ee
We can also compute the corrections to the ground state
energy due to the interaction $\frac{1}{4}C_0(\psi^\dagger
\psi)^2$. The first term is a two-loop diagram with one 
insertion of $C_0$, see Fig.~\ref{fig_fl}. We have
\be 
\label{e1}
{\cal E}_1 = C_0\left(\frac{p_F^3}{6\pi^2}\right)^2.
\ee
We should note that equ.~(\ref{e1}) contains two 
possible contractions, usually called the direct and
the exchange term. If the fermions have spin $s$ and
degeneracy $g=(2s+1)$ then equ.~(\ref{e1}) has to be 
multiplied by a factor $g(g-1)/2$. We also note that 
the sum of the first two terms in the energy density 
can be written 
as 
\be
\label{e_pfa}
{\cal E} = \rho\frac{p_F^2}{2m}\left(
\frac{3}{5} + \frac{2}{3\pi}(p_Fa)+\ldots  \right),
\ee
which shows that the $C_0$ term is the first term in 
an expansion in $p_Fa$, suitable for a dilute, weakly
interacting, Fermi gas. The expansion in $(p_Fa)$ was 
carried out to order $(p_Fa)^2$ by Huang, Lee and Yang 
\cite{Lee:1957,Huang:1957}. Since then, the accuracy
was pushed to $O((p_Fa)^4\log(p_Fa))$ \cite{Fetter:1971}, 
see \cite{Hammer:2000xg} for a modern perspective. The
effective lagrangian can also be used to study many 
other properties of the system, such as corrections
to the fermion propagator. Near the Fermi surface the 
propagator can be written as
\be
\label{s_qp}
 S_{\alpha\beta} = \frac{Z\delta_{\alpha\beta}}
 {p_0-v_F(|\vec{p}|-p_F)
   +i\delta{\rm sgn}(|\vec{p}|-p_F)},
\ee
where $Z$ is the wave function renormalization and 
$v_F=p_F/m^*$ is the Fermi velocity. $Z$ and $m^*$ 
can be worked out order by order in $(p_Fa)$, see
\cite{Abrikosov:1963,Platter:2002yr}. The main observation
is that the structure of the propagator is unchanged
even if interactions are taken into account. The 
low energy excitations are quasi-particles and holes, 
and near the Fermi surface the lifetime of a quasi-particle
is infinite. This is the basis of Fermi liquid theory, 
see Sec.~\ref{sec_fl}. We should note, however, that
for nuclear systems the $(p_Fa)$ expansion is not particularly 
useful since the nucleon-nucleon scattering length is very 
large. Equ.~(\ref{e_pfa}) is of interest for trapped dilute 
Fermi gases.

\begin{figure}
\includegraphics[width=15.0cm]{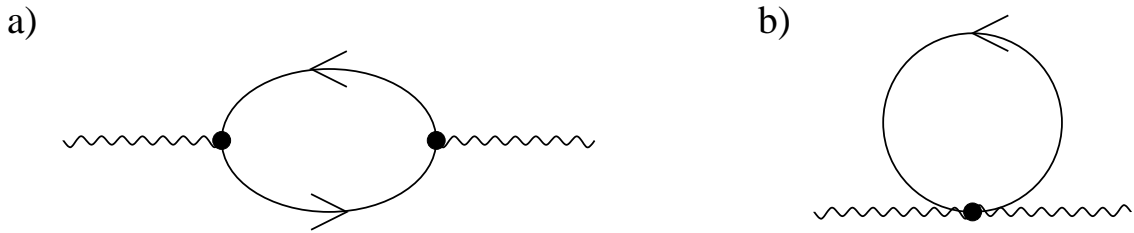}
\caption{\label{fig_screen}
Leading order Feynman diagrams that contribute to the 
photon polarization function in a non-relativistic Fermi 
liquid. The tadpole diagram shown in the right panel 
only appears in the spatial part of the polarization 
tensor.}
\end{figure}

\subsection{Screening and damping}
\label{sec_screen}

An important aspect of the dilute Fermi gas is the 
response to an external electromagnetic field. As
a simple example we will consider the case of an 
a static electric field. The coupling of the gauge
field is given by $eA_0\psi^\dagger \psi$. The 
medium correction to the photon propagator is 
determined by the polarization function
\be
\Pi_{00}(q) = e^2 \int d^4x\, e^{-iqx}
 \langle \psi^\dagger \psi(0)\psi^\dagger \psi(x)
\rangle .
\ee
The one-loop contribution is given by
\be
\Pi_{00}(q) = e^2 \int\frac{d^4p}{(2\pi)^4}
 \frac{1}{q_0+p_0-\epsilon_{p+q}+
        i\delta{\rm sgn}(\epsilon_{p+q})}
 \frac{1}{p_0-\epsilon_{p}+
        i\delta{\rm sgn}(\epsilon_{p})}.
\ee
Performing the $p_0$ integration by picking up the 
pole we find
\be
 \Pi_{00}(q) = e^2\int\frac{d^3p}{(2\pi)^3} 
\frac{n_{p+q}-n_p}{E_{p+q}-E_p},
\ee
where we have introduced the Fermi distribution function
$n_p=\Theta(p_F-p)$. We observe that in the limit 
$\vec{q}\to 0$ the polarization function only 
receives contributions from particle-hole pairs
that are closer and closer to the Fermi surface.
On the other hand, the energy denominator diverges
in this limit because the photon can excite 
particle-hole pairs with arbitrarily small energy. 
As a result we get a finite contribution
\be
\label{pi00}
\Pi_{00}(q_0=0,\vec{q}\to 0) = 
e^2\int\frac{d^3p}{(2\pi)^3} \frac{\partial n_p}{\partial E_p}
 = e^2\frac{p_Fm}{2\pi^2},
\ee
which is proportional to the density of states on the 
Fermi surface. Equ.~(\ref{pi00}) implies that the static 
photon propagator in the limit $\vec{q}\to 0$ is modified 
according to $1/\vec{q}^{\,2} \to 1/(\vec{q}^{\,2}+m_D^2)$, 
where 
\be 
\label{m_D}
m_D^2=e^2 \left( \frac{p_Fm}{2\pi^2}\right)
\ee
is called the Debye mass. The factor $N=(p_Fm)/(2\pi^2)$ 
is equal to the density of states on the Fermi surface.
In a relativistic theory we find the same result as 
in equ.~(\ref{m_D}) with the density of states replaced
by the correct relativistic expression $N=(p_FE_F)/(2\pi^2)$.
The Coulomb potential is modified as
\be 
V(r)= - e\frac{e^{-r/r_D}}{r},
\ee
where $r_D=1/m_D$ is called the Debye screening length.
The physics of screening is very easy to understand.
A test charge can polarize virtual particle-hole pairs
that act to shield the charge. 

 In the same fashion we can study the response to an 
external vector potential $\vec{A}$. The coupling of 
a non-relativistic fermion to the vector potential is 
determined in the usual way by replacing $\vec{p}\to
\vec{p}+e\vec{A}$. Since the kinetic energy operator 
is quadratic in the momentum we find a linear and a
quadratic coupling of the vector potential. The one-loop
diagrams that contribute to the polarization tensor 
are shown  in Fig.~\ref{fig_screen}. In the limit 
of small external momenta we find
\be
 \Pi_{ij}(q) = e^2 m_D^2
\int\frac{d\Omega}{4\pi}\left\{
 v_i v_j \frac{vk (\hat{q}\cdot\hat{p})}
     {q_0-vk(\hat{q}\cdot\hat{p})}
  -\frac{1}{3}v^2\delta_{ij} \right\},
\ee
where $\vec{v}=\vec{p}/m$ is the Fermi velocity. In the 
limit $q_0\!=\!0$ the polarization tensor vanishes. There 
is no screening of static magnetic fields. For non-zero 
$q_0$ the trace of the polarization tensor is given by
\be
\label{pi_nr_ii}
\Pi_{ii}(q) = m_D^2\frac{vq_0}{2q}
  \log\left( \frac{q_0-vq}{q_0+vq}\right).
\ee
The result equ.~(\ref{pi_nr_ii}) has an imaginary 
part if $vq>q_0$. This phenomenon is known as Landau 
damping. The photon is loosing energy to electrons 
in the Fermi liquid. For a discussion in the context 
of kinetic theory we refer the reader to \cite{Landau:kin}.

\begin{figure}
\includegraphics[width=13.0cm]{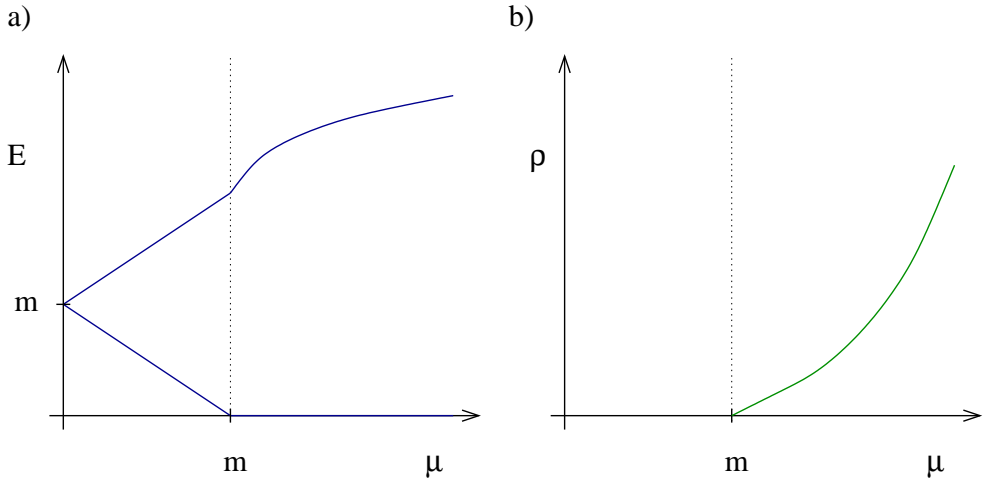}
\caption{\label{fig_bose}
Spectrum and charge density for a charged scalar field
as a function of the chemical potential.}
\end{figure}

\section{Bose condensation}
\label{sec_bose}

 In this section we introduce some general
features of bosonic systems at finite density. We 
will consider a charged relativistic boson described
by the Lagrange density
\be
{\cal L}_0 = (\partial^\mu \phi^*)(\partial_\mu \phi)
-m^2\phi^*\phi - \lambda(\phi^*\phi)^2.
\ee
Note that $\lambda$ has to be positive in order for
the theory to be well defined. This corresponds to
a repulsive interaction between the bosons. The lagrangian 
has a $U(1)$ symmetry $\phi\to e^{-i\varphi}\phi$. The 
corresponding conserved charge is 
\be 
Q = \int d^3x \, i\left( \phi^*\partial_0\phi
 - \phi\partial_0\phi^* \right).
\ee
Note that the charge density $\rho$ contains not only 
the field $\phi$ but also the canonically conjugate momentum 
$\partial_0\phi$. This means that the chemical potential 
modifies the integration over the canonical momenta in 
the path integral representation of the partition function. 
The resulting Minkowski space path integral is
\cite{Kapusta:1989,LeBellac:1996}
\be 
\label{Z_bos}
Z = \int D\phi D\phi^* \exp\left(i\int d^4x {\cal L}\right) ,
\ee
with 
\be 
\label{l_bos}
{\cal L} = (\partial_0+i\mu) \phi^*(\partial_0-i\mu) \phi
  -(\vec\nabla\phi^*)(\vec\nabla\phi)
  -m^2\phi^*\phi - \lambda(\phi^*\phi)^2.
\ee
There is a simple argument that fixes the 
form of the lagrangian equ.~(\ref{l_bos}). The argument
is based on the observation that we can promote the 
global $U(1)$ symmetry to a local symmetry by adding 
a $U(1)$ gauge field to the lagrangian. The charge density 
is obtained by varying the effective action with respect 
to the gauge potential. This implies that the chemical 
potential has to enter the lagrangian like the time 
component of a gauge field. 

 We can study the effect of a chemical potential in the 
mean field approximation. The classical effective potential 
for the field $\phi$ is given by 
\be 
 V(\phi) = (m^2-\mu^2) (\phi^*\phi) + \lambda 
  (\phi^*\phi)^2 .
\ee
For $\mu>m$ the quadratic term is positive and the minimum 
of the effective potential is at $\langle \phi\rangle = 0$.
For $\mu>m$ the origin is unstable and
\be
\langle \phi \rangle^2  =\frac{\mu^2-m^2}{2\lambda}.
\ee
This state is a Bose condensate. The charge density is 
\be 
\label{rho_bose}
\rho = \frac{\mu}{\lambda}(\mu^2-m^2). 
\ee
In a non-interacting Bose gas the chemical potential
cannot be larger than the mass of the boson. In 
our case, repulsive interactions limit the growth of 
the density and the chemical potential can take any 
value. We can also compute the spectrum as a function
of the chemical potential. We write $\phi=\langle \phi
\rangle +\chi_1+i\chi_2$ and expand the effective 
action to second order in $\chi_{1,2}$. For $\mu<m$ we 
find two modes with energies $E(\vec{p}\!=\!0)=m\pm\mu$.
Bose condensation sets in when the lower mode reaches
zero energy. Above the onset of Bose condensation we 
find
\be 
E_1(\vec{p}\!=\!0)=0, \hspace{1cm}
E_2(\vec{p}\!=\!0)=\sqrt{6\mu^2-2m^2}.
\ee
Bose condensation breaks the $U(1)$ symmetry spontaneously
and the spectrum contains one Goldstone boson. It is also
interesting to study the dispersion relation of the 
Goldstone mode in more detail. For small momenta we find
\be
E_1(\vec{p}) = \sqrt{1-\frac{2\mu^2}{3\mu^2-m^2}}
 |\vec{p}| +\ldots .
\ee
This shows that at the phase transition point the 
velocity of the Goldstone mode is zero. Far away from
the transition the velocity approaches $v=c/\sqrt{3}$.
Bose condensates have many remarkable properties, most 
notably the fact that they can flow without viscosity. 
These properties can be derived from the effective 
action for the Goldstone mode. It was shown, in particular,
that this effective action is equivalent to superfluid 
hydrodynamics \cite{popov:1987,Son:2002zn}.

\section{Superconductivity}
\label{sec_sc}
\subsection{BCS instability}
\label{sec_bcs}

\begin{figure}
\includegraphics[width=15.0cm]{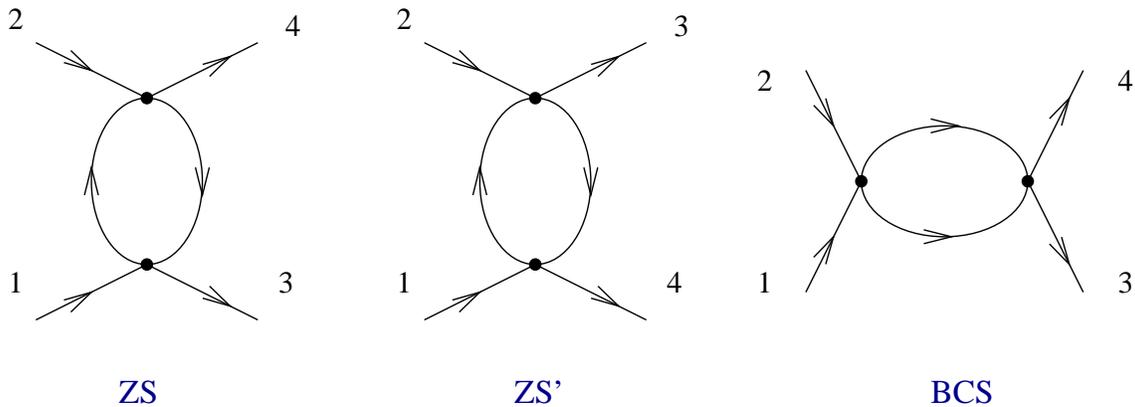}
\caption{\label{fig_bcs}
Second order diagrams that contribute to particle-particle
scattering. The three diagrams are known as ZS (zero sound),
ZS' and BCS (Bardeen-Cooper-Schrieffer) contribution.}
\end{figure}

 One of the most remarkable phenomena that take place in 
many body systems is superconductivity. Superconductivity 
is related to an instability of the Fermi surface in the 
presence of attractive interactions between fermions. Let 
us consider fermion-fermion scattering in the simple
model introduced in Sect.~\ref{sec_fl}. At leading 
order the scattering amplitude is given by
\be
\label{pp_0}
\Gamma_{\alpha\beta\gamma\delta}(p_1,p_2,p_3,p_4) = 
C_0 \left( \delta_{\alpha\gamma}\delta_{\beta\delta}
 - \delta_{\alpha\delta}\delta_{\beta\gamma} \right).
\ee
At next-to-leading order we find the corrections shown 
in Fig.~\ref{fig_bcs}. A detailed discussion of the role 
of these corrections can be found in 
\cite{Abrikosov:1963,Shankar:1993pf,Polchinski:1992ed}.
The BCS diagram is special, because in the case of a
spherical Fermi surface it can lead to an instability 
in weak coupling. The main point is that if the 
incoming momenta satisfy $\vec{p}_1\simeq -\vec{p}_2$
then there are no kinematic restrictions on the loop
momenta. As a consequence, all back-to-back pairs can
mix and there is an instability even in weak coupling. 

 For $\vec{p}_1= -\vec{p}_2$ and $E_1=E_2=E$ the BCS 
diagram is given by
\be
\label{diag_bcs}
\Gamma_{\alpha\beta\gamma\delta} = 
C_0^2 \left( \delta_{\alpha\gamma}\delta_{\beta\delta}
 - \delta_{\alpha\delta}\delta_{\beta\gamma} \right)
\int \frac{d^4q}{(2\pi)^4} 
 \frac{1}{E+q_0-\epsilon_q+i\delta{\rm sgn}(\epsilon_q)}
 \frac{1}{E-q_0-\epsilon_q+i\delta{\rm sgn}(\epsilon_q)}.
\ee
The loop integral has an infrared divergence near 
the Fermi surface as $E\to 0$. The scattering amplitude 
is proportional to 
\be 
\label{cor_bcs}
\Gamma_{\alpha\beta\gamma\delta} =
\left( \delta_{\alpha\gamma}\delta_{\beta\delta}
 - \delta_{\alpha\delta}\delta_{\beta\gamma} \right)
\left\{
C_0 - C_0^2\left(\frac{p_Fm}{2\pi^2}\right)
 \log\left(\frac{E_0}{E}\right) \right\},
\ee
where $E_0$ is an ultraviolet cutoff. Equ.~(\ref{cor_bcs})
can be interpreted as an effective energy dependent 
coupling that satisfies the renormalization group 
equation \cite{Shankar:1993pf,Polchinski:1992ed}
\be 
\label{rge_bcs}
 E\frac{dC_0}{dE} = C_0^2 \left(\frac{p_Fm}{2\pi^2}\right),
\ee
with the solution
\be
\label{rge_sol}
C_0(E) =\frac{C_0(E_0)}{1+NC_0(E_0)\log(E_0/E)},
\ee
where $N=(p_Fm)/(2\pi^2)$ is the density of states. 
Equ.~(\ref{rge_sol}) shows that there are two possible
scenarios. If the initial coupling is repulsive, $C_0
(E_0)>0$, then the renormalization group evolution will 
drive the effective coupling to zero and the Fermi liquid 
is stable. If, on the other hand, the initial coupling is 
attractive, $C_0(E_0)<0$, then the effective coupling 
grows and reaches a Landau pole at 
\be 
\label{E_lp}
 E_{\it crit} \sim E_0 
    \exp\left(-\frac{1}{N|C_0(E_0)|}\right).
\ee
At the Landau pole the Fermi liquid description has to 
break down. The renormalization group equation does not
determine what happens at this point, but it seems 
natural to assume that the strong attractive interaction
will lead to the formation of a fermion pair condensate. 
The fermion condensate $\langle\epsilon^{\alpha\beta}
\psi_\alpha\psi_\beta\rangle$ signals the breakdown 
of the $U(1)$ symmetry and leads to a gap $\Delta$ in 
the single particle spectrum. 

 The scale of the gap is determined by the position 
of the Landau pole, $\Delta\sim E_{\it crit}$. A more 
quantitative estimate of the gap can be obtained in the 
mean field approximation. In the path integral formulation 
the mean field approximation is most easily introduced
using the Hubbard-Stratonovich trick. For this purpose
we first rewrite the four-fermion interaction as
\be 
\label{4f_fierz}
\frac{C_0}{2}(\psi^\dagger\psi)^2  = 
\frac{C_0}{4} \left\{
 (\psi^\dagger\sigma_2\psi^\dagger)
 (\psi\sigma_2\psi) 
+(\psi^\dagger\sigma_2\vec{\sigma}\psi^\dagger)
 (\psi\vec{\sigma}\sigma_2\psi)\right\},
\ee 
where we have used the Fierz identity $2\delta^{\alpha
\beta}\delta^{\gamma\rho} = \delta^{\alpha\rho}
\delta^{\gamma\beta}+(\vec{\sigma})^{\alpha\rho}
(\vec{\sigma})^{\gamma\beta}$. Note that the second 
term in equ.~(\ref{4f_fierz}) vanishes because 
$(\sigma_2\vec{\sigma})$ is a symmetric matrix. We 
now introduce a factor of unity into the path integral
\be
1 = \frac{1}{Z_\Delta}\int D\Delta 
\exp\left(\frac{\Delta^*\Delta}{C_0}\right),
\ee
where we assume that $C_0<0$. We can eliminate the 
four-fermion term in the lagrangian by a shift in the 
integration variable $\Delta$. The action is now 
quadratic in the fermion fields, but it involves 
a Majorana mass term $\psi\sigma_2\Delta \psi+h.c$. The
Majorana mass terms can be handled using the Nambu-Gorkov 
method. We introduce the bispinor $\Psi=(\psi,\psi^\dagger
\sigma_2)$ and write the fermionic action as
\be
\label{s_ng}
{\cal S} = \frac{1}{2}\int\frac{d^4p}{(2\pi)^4}
 \Psi^\dagger
 \left(\begin{array}{cc}
     p_0-\epsilon_p  & \Delta \\
     \Delta^* & p_0+\epsilon_p
 \end{array}\right) \Psi.
\ee
Since the fermion action is quadratic we can integrate
the fermion out and obtain the effective lagrangian
\be
\label{s_ng_eff}
{\cal L}= \frac{1}{2}{\rm Tr}\left[\log\left(
 G_0^{-1}G\right)\right]+\frac{1}{C_0}|\Delta|^2,
\ee
where $G$ is the fermion propagator
\be
\label{ng_prop}
 G(p) = \frac{1}{p_0^2-\epsilon_p^2-|\Delta|^2}
 \left(\begin{array}{cc}
     p_0+\epsilon_p  & \Delta^* \\
     \Delta & p_0-\epsilon_p
 \end{array}\right).
\ee
The diagonal and off-diagonal components of $G(p)$ are 
sometimes referred to as normal and anomalous propagators. 
Note that we have not yet made any approximation. We have 
converted the fermionic path integral to a bosonic one, albeit 
with a very non-local action. The mean field approximation 
corresponds to evaluating the bosonic path integral using 
the saddle point method. Physically, this approximation
means that the order parameter does not fluctuate. 
Formally, the mean field approximation can be 
justified in the large $N$ limit, where $N$ is the
number of fermion fields. The saddle point equation
for $\Delta$ gives the gap equation
\be
\Delta = |C_0|\int\frac{d^4p}{(2\pi)^4} 
 \frac{\Delta}{p_0^2-\epsilon^2_p-\Delta^2}.
\ee
Performing the $p_0$ integration we find
\be
\label{4f_gap}
1 = \frac{|C_0|}{2}\int\frac{d^3p}{(2\pi)^3} 
 \frac{1}{\sqrt{\epsilon^2_p+\Delta^2}}.
\ee
Since $\epsilon_p=E_p-\mu$ the integral in equ.~(\ref{4f_gap}) 
has an infrared divergence on the Fermi surface $|\vec{p}|
\sim p_F$. As a result, the gap equation has a non-trivial 
solution even if the coupling is arbitrarily small. The 
magnitude of the gap is $\Delta\sim \Lambda \exp(-1/(|C_0|N))$ 
where $\Lambda$ is a cutoff that regularizes the integral 
in equ.~(\ref{4f_gap}) in the ultraviolet. If we treat
equ.~(\ref{l_4f}) as a low energy effective field theory 
we should be able to eliminate the unphysical dependence 
of the gap on the ultraviolet cutoff, and express the gap 
in terms of a physical observable. At low density, this 
can be achieved by observing that the gap equation has 
the same UV behavior as the Lipmann-Schwinger equation 
that determines the scattering length at zero density
\be
\label{bubble}
\frac{mC_0}{4\pi a} - 1 = \frac{C_0}{2} 
\int\frac{d^3p}{(2\pi)^3}\frac{1}{E_P}.
\ee
Combining equs.~(\ref{4f_gap}) and (\ref{bubble}) we
can derive an UV finite gap equation that depends only
on the scattering length, 
\be
-\frac{m}{4\pi a} = 
\frac{1}{2}\int\frac{d^3p}{(2\pi)^3} \Big\{
 \frac{1}{\sqrt{\epsilon^2_p+\Delta^2}}
 -\frac{1}{E_p}\Big\}.
\ee 
A careful analysis gives \cite{Papenbrock:1998wb,Khodel:1996}
\be
\label{gap_lowd}
\Delta = \frac{8E_f}{e^2}\exp\left(-\frac{\pi}{2p_F|a|}\right).
\ee
For neutron matter the scattering length is large, 
$a=-18.8$ fm, and equ.~(\ref{gap_lowd}) is not very
useful, except at very small density. Calculations 
based on potential models give gaps on the order of 
2 MeV at nuclear matter density.

In the limit of very high density we can eliminate 
the cutoff dependence using a method introduced by 
Weinberg \cite{Weinberg:1994}. Weinberg defines a
renormalized effective potential and shows that the 
renormalization scale dependence of the effective
potential is canceled by the scale dependence of the 
coupling. The effective coupling satisfies the 
renormalization group equ.~(\ref{rge_bcs}). The 
gap is determined by the effective coupling at the 
energy scale $E_0$. In practice, this would typically
be the energy scale at which the four-fermion 
interaction is matched against a more microscopic 
description in terms of meson (nuclear physics)
or phonon exchange (condensed matter physics).

\subsection{Fermi liquid, revisited}
\label{sec_flr}

 Our discussion of Fermi liquids in Sect.~\ref{sec_fl} and 
in the previous section was based on the simple model 
defined in equ.~(\ref{l_4f}). In this section we shall
briefly discuss the structure of fermionic many-body 
systems in the case of more general interactions. We 
will restrict ourselves to systems that can be described
in terms of purely fermionic actions, with all other 
degrees of freedom integrated out. For more details 
we refer the reader to \cite{Shankar:1993pf}.

\begin{figure}
\includegraphics[width=13.0cm]{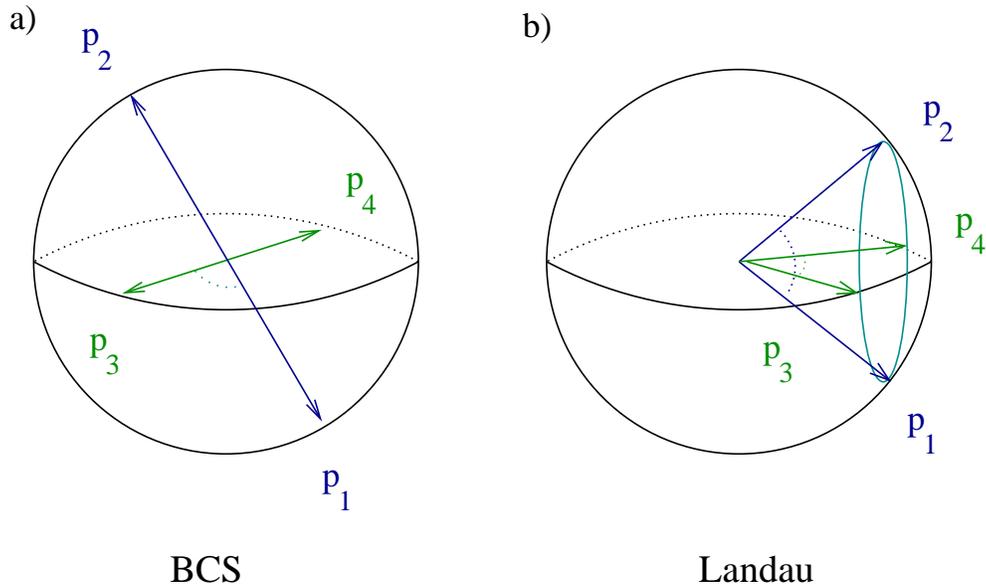}
\caption{\label{fig_fskin}
Kinematic configurations for fermion-fermion scattering 
that correspond to four-fermion operators that are marginal
at tree level. The left panel shows BCS (back-to-back) 
scattering and the right panel shows forward scattering.}
\end{figure}

 We can view the model defined by equ.~(\ref{l_4f}) as 
an example of an effective field theory, valid for momenta
close to the Fermi surface. In order to construct an effective 
field theory we have to write all possible interactions 
that are allowed by the symmetries of the theory. The 
effective action of rotationally invariant, non-relativistic
Fermi system is given by
\be 
\label{seff_fs}
S = \int \frac{d^4p}{(2\pi)^4}
 \psi(p)^\dagger\left(p_0-v_Fl_p\right)\psi(p) 
 + \frac{1}{4}\left[ \prod_{i=1}^{4}\int
     \frac{d^4p_i}{(2\pi)^4}\right]
     \psi^\dagger(p_4)\psi^\dagger(p_3)
     \psi(p_2)\psi(p_1) U(p_4,p_3,p_2,p_1),
\ee
where $v_F=\partial\epsilon_p/(\partial p)$ and $l_p
=|\vec{p}|-p_F$. We have suppressed the spin indices of
the potential $U$. The power counting for the effective 
theory can be established by studying the scaling behavior 
of all allowed operators under transformations of the type 
$l_p\to sl_p$ that scale the momenta towards the Fermi 
surface. Writing $\epsilon_p=v_Fl_p+O(l_p^2)$ we see
that as $s\to 0$ only the Fermi velocity survives,
the detailed form of the dispersion relation is irrelevant.
Using this method we can also see that with the exception
of special kinematic regimes the four-fermion interaction 
is irrelevant. We already saw that one exception is 
provided by the BCS interaction
\be
U(-\hat{p}_3,\hat{p}_3,-\hat{p}_1,\hat{p}_1) 
= V(\hat{p}_1\cdot\hat{p}_3) 
= \sum_l V_l P_l(\hat{p}_1\cdot\hat{p}_3),
\ee
where $P_l(x)$ are Legendre polynomials. 
At tree level $V(x)$ is a marginal operator, that means
it is invariant under rescaling the momenta towards the
Fermi surface. This changes at one-loop level. If any 
of the couplings $V_l$ is attractive then this coupling
will grow according to the renormalization group 
equ.~(\ref{rge_bcs}) and eventually reach a Landau pole. 
If there is more than one attractive coupling $V_l$ 
then the ground state is determined by which coupling
reaches its Landau pole first. If all $V_l$ are repulsive
then the BCS potential becomes irrelevant as the 
evolution approaches the Fermi surface. 

 In this case there is another kinematic regime that 
becomes important. We can take any two momenta on the 
Fermi surface, not necessarily back-to-back, and find
an allowed final state. Energy and momentum conservation
implies that $\hat{p}_1\cdot\hat{p}_2=\hat{p}_3\cdot
\hat{p}_4$. In two-dimensions this would restrict 
the scattering to be either forward or exchange, but
in three dimensions there is a circle of allowed 
final states parametrized by the angle $\phi_{12,34}$ 
between the planes spanned by the incoming and outgoing
momenta, see Fig.~\ref{fig_fskin}. The interaction is
\be
\left. U(\hat{p}_4,\hat{p}_3,\hat{p}_2,\hat{p}_1) 
\right|_{\hat{p}_1\cdot\hat{p}_2=\hat{p}_3\cdot\hat{p}_4}
= F(\hat{p}_1\cdot\hat{p}_2,\phi_{12,34})
\ee
The function $F(x,0)$ is called the Landau function 
and its Legendre coefficients are referred to as Landau
parameters. The Landau parameters remain marginal at 
one-loop order. A many body system characterized by 
$v_F$ and $F_l$ is called a Landau Fermi liquid
\cite{Pines:1966,Baym:1991}. A Landau liquid behaves
qualitatively like a dilute, weakly interacting, Fermi 
liquid, even if the interaction is not weak and the 
system is not dilute. In particular, the excitations 
of a Landau liquid are quasi-particles and holes. The
Landau parameters encode the quasi-particle interaction
and can be used to compute observables like the equation
of state and the response to external fields. 

  One can show that operators involving more than 
four fermion fields are irrelevant near the Fermi 
surface. This does not imply that $n>4$ fermion 
operators play no role at all. For example, if one 
of the $V_l$ has a Landau pole at energy $E_0$ 
then the six fermion interaction still has a finite 
coupling at this scale and will cause observable 
effects, see Sec.~\ref{sec_inst} for an example. 
Also, just because the $V_l$ are the only operators
that cause instabilities in weak coupling does not
imply that other operators cannot have instabilities
in strong coupling. For example, dilute nuclear 
matter may have a phase characterized by alpha 
particle condensation rather than superconductivity.

\subsection{Landau-Ginzburg theory}
\label{sec_lg}

 In this section we shall study the properties of a 
superconductor in more detail. For definiteness, we will
consider a system of electrons coupled to a $U(1)$ gauge 
field $A_\mu$. The order parameter $\Phi=\langle \epsilon^{\alpha
\beta}\psi_\alpha\psi_\beta\rangle$ breaks $U(1)$ invariance. 
Consider a gauge transformation 
\be 
A_\mu\to A_\mu +\partial_\mu\Lambda .
\ee
The order parameter transforms as
\be 
\Phi \to \exp(2ie\Lambda)\Phi.
\ee
The breaking of gauge invariance is responsible for most of the 
unusual properties of superconductors \cite{Anderson:1984,Weinberg:1995}.
This can be seen by constructing the low energy effective action 
of a superconductor. For this purpose we write the order parameter
in terms of its modulus and phase
\be 
\Phi(x) = \exp(2ie\phi(x)) \tilde\Phi(x).
\ee
The field $\phi$ corresponds to the Goldstone mode. Under a gauge
transformation $\phi(x)\to\phi(x)+\Lambda(x)$. 
Gauge invariance restricts the form of the effective Lagrange
function as 
\be 
\label{L_sc}
 L = -\frac{1}{4}\int d^3x\, F_{\mu\nu}F_{\mu\nu}
 + L_s (A_\mu-\partial_\mu\phi).
\ee
There is a large amount of information we can extract even 
without knowing the explicit form of $L_s$. Stability implies
that $A_\mu=\partial_\mu\phi$ corresponds to a minimum of the 
energy. This means that up to boundary effects the gauge 
potential is a total divergence and that the magnetic field
has to vanish. This phenomenon is known as the Meissner
effect. 

 Equ.~(\ref{L_sc}) also implies that a superconductor
has zero resistance. The equations of motion relate
the time dependence of the Goldstone boson field to 
the potential, 
\be
\label{phidot}
\dot\phi(x)=-V(x).
\ee 
The electric current is related to the gradient of the 
Goldstone boson field. Equ.~(\ref{phidot}) shows that the 
time dependence of the current is proportional to the 
gradient of the potential. In order to have a static 
current the gradient of the potential has to be constant
throughout the sample, and the resistance is zero. 

 In order to study the properties of a superconductor in 
more detail we have to specify $L_s$. For this purpose we 
assume that the system is time-independent, that the spatial 
gradients are small, and that the order parameter is small. 
In this case we can write
\be 
\label{l_lg}
L_s = \int d^3x\, \left\{
-\frac{1}{2}\left|\left(\nabla-2ie\vec{A}\right)\Phi\right|^2
 +\frac{1}{2}m^2_H\left(\Phi^*\Phi\right)^2
 -\frac{1}{4}g\left(\Phi^*\Phi\right)^4 + \ldots \right\},
\ee
where $m_H$ and $g$ are unknown parameters that depend
on the temperature. Equ.~(\ref{l_lg}) is known as the 
Landau-Ginzburg effective action. Strictly speaking, the 
assumption that the order parameter is small can only be 
justified in the vicinity of a second order phase transition. 
Nevertheless, the Landau-Ginzburg description is instructive 
even in the regime where $t=(T-T_c)/T_c$ is not small. It is 
useful to decompose $\Phi=\rho\exp(2ie\phi)$. For constant
fields the effective potential, 
\be
\label{v_lg}
V(\rho)=-\frac{1}{2}m_H^2\rho^2 +\frac{1}{4}g\rho^4 ,
\ee
is independent of $\phi$. The minimum is at $\rho_0^2=m_H^2/g$ 
and the energy density at the minimum is given by ${\cal E}= 
-m_H^4/(4g)$. This shows that the two parameters $m_H$ and
$g$ can be related to the expectation value of $\Phi$ and
the condensation energy. We also observe that the phase
transition is characterized by $m_H(T_c)=0$. 

 In terms of $\phi$ and $\rho$ the Landau-Ginzburg action 
is given by
\be 
L_s = \int d^3x\, \left\{
-2e^2\rho^2 \left(\vec\nabla\phi-\vec{A}\right)^2
 +\frac{1}{2}m_H^2\rho^2 -\frac{1}{4}g\rho^4
 -\frac{1}{2}\left(\nabla\rho\right)^2
\right\}.
\ee
The equations of motion for $\vec{A}$ and $\rho$ 
are given by
\bea 
\label{b_lg}
\vec\nabla\times \vec{B} &=& 
 4e^2\rho^2 \left(\nabla\phi -\vec{A}\right), \\
\label{rho_lg}
 \nabla^2 \rho &=& 
 -m_H^2\rho^2 + g\rho^3 + 4e^2 \rho 
 \left( \vec\nabla\phi-\vec{A}\right) .
\eea
Equ.~(\ref{b_lg}) implies that $\nabla^2\vec{B} = 
-4e^2\rho^2\vec{B}$. This means that an external
magnetic field $\vec{B}$ decays over a characteristic
distance $\lambda=1/(2e\rho)$. Equ.~(\ref{rho_lg}) 
gives $\nabla^2\rho = -m_H^2\rho+\ldots$. As a consequence,
variations in the order parameter relax over a length 
scale given by $\xi=1/m_H$. The two parameters $\lambda$
and $\xi$ are known as the penetration depth and the 
coherence length. 

 The relative size of $\lambda$ and $\xi$ has important 
consequences for the properties of superconductors. In 
a type II superconductor $\xi<\lambda$. In this case 
magnetic flux can penetrate the system in the form of 
vortex lines. At the core of a vortex the order parameter
vanishes, $\rho=0$. In a type II material the core is 
much smaller than the region over which the magnetic
field goes to zero. The magnetic flux is given by
\be
\int_A\vec{B}\cdot\vec{S} =
\oint_{\partial A} \vec{A}\cdot d\vec{l} = 
\oint_{\partial A} \vec{\nabla}\phi \cdot d\vec{l} =
\frac{n\pi\hbar}{e} ,
\ee
and quantized in units of $\pi\hbar/e$. In a type II
superconductor magnetic vortices repel each other and 
form a regular lattice known as the Abrikosov lattice. 
In a type I material, on the other hand, vortices are 
not stable and magnetic fields can only penetrate
the sample if superconductivity is destroyed. 

 The Landau-Ginzburg description shows that there is 
no qualitative difference between superconductivity 
and Bose condensation of charged bosons. Indeed, we 
may think of superconductivity as Bose condensation
of Cooper pairs. While this is qualitatively correct,
there is an important quantitative difference between
a BCS superconductor and a dilute Bose condensate of 
composite bosons. In a BCS superconductor the coherence
length $\xi$, which is a measure of the size of the 
Cooper pairs, is much larger than the average inter-particle
spacing $p_F^{-1}$. Also, the pair correlation essentially 
disappears above the critical temperature. In a dilute 
Bose condensate, on the other hand, the size of the bosons 
is much smaller than the typical distance between them. 
The bosons are tightly bound and do not dissolve at 
$T_c$. Nevertheless, since there is no qualitative difference 
between Bose condensation and BCS superconductivity we
expect to find systems that show a crossover from one
kind of behavior to the other. We will discuss
an example in Sect.~\ref{sec_nc2}.

\section{QCD and symmetries}
\label{sec_qcd}

 Before we discuss QCD at finite baryon density we would
like to provide a quick reminder on QCD and the symmetries
of QCD. The elementary degrees of freedom are quark fields 
$\psi^a_{\alpha,f}$ and gluons $A_\mu^a$. Here, $a$ is 
color index that transforms in the fundamental representation
for fermions and in the adjoint representation for gluons. 
Also, $f$ labels the quark flavors $u,d,s,c,b,t$. In practice, 
we will focus on the three light flavors up, down and strange.
The QCD lagrangian is 
\be
\label{l_qcd}
 {\cal L } = \sum_f^{N_f} \bar{\psi}_f ( i\Dslash - m_f) \psi_f
  - \frac{1}{4} G_{\mu\nu}^a G_{\mu\nu}^a,
\ee
where the field strength tensor is defined by 
\be
 G_{\mu\nu}^a = \partial_\mu A_\nu^a - \partial_\nu A_\mu^a
  + gf^{abc} A_\mu^b A_\nu^c,
\ee
and the covariant derivative acting on quark fields is 
\be
 i\Dslash \psi = \gamma^\mu \left(
 i\partial_\mu + g A_\mu^a \frac{\lambda^a}{2}\right) \psi.
\ee
QCD has a number of remarkable properties. Most remarkably, 
even though QCD accounts for the rich phenomenology of hadronic
and nuclear physics, it is an essentially parameter free 
theory. To first approximation, the masses of the light
quarks $u,d,s$ are too small to be important, while the masses 
of the heavy quarks $c,b,t$ are too heavy. If we set 
the masses of the light quarks to zero and take the masses
of the heavy quarks to be infinite then the only parameter 
in the QCD lagrangian is the coupling constant, $g$. Once
quantum corrections are taken into account $g$ becomes 
a function of the scale at which it is measured. If the 
scale is large then the coupling is small, but in the 
infrared the coupling becomes large. This is the famous
phenomenon of asymptotic freedom. Since the coupling 
depends on the scale the dimensionless parameter $g$ is 
traded for a dimensionful scale parameter $\Lambda_{QCD}$. 
In essence, $\Lambda_{QCD}$ is the scale at which the 
coupling becomes large. 

 Since $\Lambda_{QCD}$ is the only dimensionful quantity
in QCD ($m_q=0$) it is not really a parameter of QCD, but
reflects our choice of units. In standard units, $\Lambda_{QCD} 
\simeq 200\,{\rm MeV} \simeq 1\,{\rm fm}^{-1}$. Note that 
hadrons indeed have sizes $r\sim\Lambda_{QCD}^{-1}$. 
However, we should also note that in practice the 
perturbative expansion in $g$ breaks down at scales
$r\sim\Lambda_{\chi SB}^{-1}\sim 0.2\,{\rm fm}\ll
\Lambda_{QCD}^{-1}$.

 Another important feature of the QCD lagrangian are 
its symmetries. First of all, the lagrangian is invariant 
under local gauge transformations $U(x)\in SU(3)_c$
\be 
\psi(x) \to U(x)\psi(x),\hspace{1cm}
A_\mu(x) \to U(x)A_\mu U^\dagger (x)
 + iU(x)\partial_\mu U^\dagger(x),
\ee
where $A_\mu= A_\mu^a(\lambda^a/2)$. In the QCD ground 
state at zero temperature and density the local color 
symmetry is confined. This implies that all excitations
are singlets under the gauge group. 

 The dynamics of QCD is completely independent of flavor. This 
implies that if the masses of the quarks are equal, $m_u=
m_d=m_s$, then the theory is invariant under arbitrary flavor 
rotations of the quark fields 
\be
 \psi_f\to V_{fg}\psi_g,
\ee
where 
$V\in SU(3)$. This is the well known flavor (isospin)
symmetry of the strong interactions. If the quark masses 
are not just equal, but equal to zero, then the flavor 
symmetry is enlarged. This can be seen by defining left 
and right-handed fields
\be
  \psi_{L,R} = \frac{1}{2} (1\pm \gamma_5) \psi .
\ee
In terms of $L/R$ fields the fermionic lagrangian is
\be
 {\cal L} =  \bar{\psi}_L (i\Dslash) \psi_L
    +\bar{\psi}_R (i\Dslash) \psi_R + 
   \bar{\psi}_L M \psi_R + \bar{\psi}_R M\psi_L ,
\ee
where $M = {\rm diag}(m_u,m_d,m_s)$. We observe that if 
quarks are massless, $m_u=m_d=m_s=0$, then there is no 
coupling between left and right handed fields. As a 
consequence, the lagrangian is invariant under independent
flavor transformations of the left and right handed fields.
\be
 \psi_{L,f}\to L_{fg}\psi_{L,g}, \hspace{1cm}
 \psi_{R,f}\to R_{fg}\psi_{R,g},
\ee
where $(L,R)\in SU(3)_L\times SU(3)_R$. In the real world, 
of course, the masses of the up, down and strange quarks 
are not zero. Nevertheless, since $m_u,m_d\ll m_s < \Lambda_{QCD}$ 
QCD has an approximate chiral symmetry. 

 In the QCD ground state at zero temperature and density 
the flavor symmetry is realized, but the chiral symmetry 
is spontaneously broken by a quark-anti-quark condensate
$\langle \bar\psi_L\psi_R +\bar\psi_R\psi_L\rangle$. As
a result, the observed hadrons can be approximately 
assigned to representations of the $SU(3)_V$ flavor group, 
but not to representations of $SU(3)_L\times SU(3)_R$. 
Nevertheless, chiral symmetry has important implications 
for the dynamics of QCD at low energy. Goldstone's theorem
implies that the breaking of $SU(3)_L\times SU(3)_R\to
SU(3)_V$ is associated with the appearance of an octet 
of (approximately) massless pseudoscalar Goldstone bosons. 
Chiral symmetry places important restrictions on the
interaction of the Goldstone bosons. These constraints
are obtained most easily from the low energy effective 
chiral lagrangian. At leading order we have
\be
\label{l_chpt}
{\cal L} = \frac{f_\pi^2}{4} {\rm Tr}\left[
 \partial_\mu\Sigma\partial^\mu\Sigma^\dagger\right] 
 +\Big[ B {\rm Tr}(M\Sigma^\dagger) + h.c. \Big]
+ \ldots, 
\ee
where $\Sigma=\exp(i\phi^a\lambda^a/f_\pi)$ is the chiral field,
$f_\pi$ is the pion decay constant and $M$ is the mass
matrix. Expanding $\Sigma$ in powers of the pion, kaon
and eta fields $\phi^a$ we can derive the leading order 
chiral perturbation theory results for Goldstone boson 
scattering and the coupling of Goldstone bosons to 
external fields. Higher order corrections originate
from loops and higher order terms in the effective 
lagrangian.

 Finally, we observe that the QCD lagrangian has two 
$U(1)$ symmetries,
\bea
U(1)_B: \hspace{1cm}& \psi_L\to e^{i\phi}\psi_L, \hspace{1cm}& 
     \psi_R\to e^{i\phi}\psi_R \\
U(1)_A: \hspace{1cm}& \psi_L\to e^{i\alpha}\psi_L,\hspace{1cm} & 
     \psi_R\to e^{-i\alpha}\psi_R .
\eea
The $U(1)_B$ symmetry is exact even if the quarks are not 
massless. Superficially, it appears that the $U(1)_A$ 
symmetry is explicitly broken by the quark masses and 
spontaneously broken by the quark condensate. However, 
there is no Goldstone boson associated with spontaneous
$U(1)_A$ breaking. The reason is that at the quantum level
the $U(1)_A$ symmetry is broken by an anomaly. The 
divergence of the $U(1)_A$ current is given by
\be 
\partial^\mu j_\mu^5 = \frac{N_f g^2}{16\pi^2}
 G^a_{\mu\nu}\tilde{G}^a_{\mu\nu},
\ee
where $\tilde{G}^a_{\mu\nu}=\epsilon_{\mu\nu\alpha\beta}
G^a_{\alpha\beta}/2$ is the dual field strength tensor.

\section{QCD at finite density}
\label{sec_dqcd}

 In the real world the quark masses are not equal and the 
only exact global symmetries of QCD are the $U(1)_f$ flavor
symmetries associated with the conservation of the number 
of up, down, and strange quarks. If we take into 
account the weak interactions then flavor is no longer 
conserved and the only exact symmetries are the $U(1)_B$ 
of baryon number and the $U(1)_Q$ of electric charge. 

 In the following we study hadronic matter at non-zero 
baryon density. We will mostly focus on systems at non-zero
baryon chemical potential but zero electron $U(1)_Q$
chemical potential. We should note that in the context
of neutron stars we are interested in situations when the 
electric charge, but not necessarily the electron chemical
potential, is zero. We will comment on the consequences
of electric charge neutrality below. Also, if the system
is in equilibrium with respect to strong, but not to weak 
interactions, then non-zero flavor chemical potentials may
come into play. 

\begin{figure}
\includegraphics[width=9.0cm]{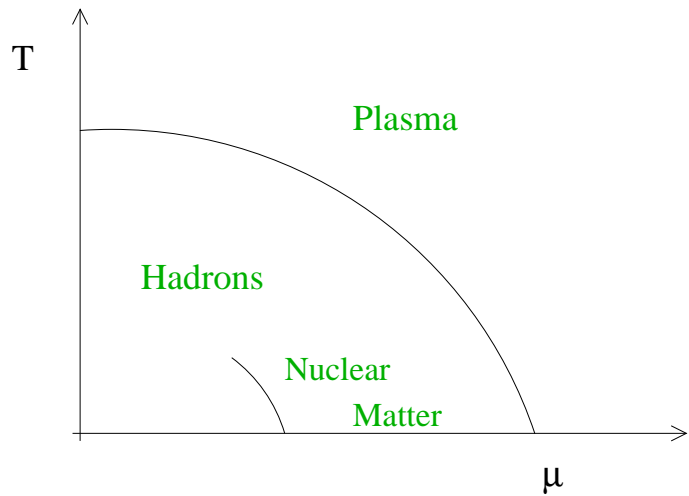}
\caption{\label{fig_phase_1}
Naive phase diagram of hadronic matter as a function of the
baryon chemical potential and temperature.}
\end{figure}

The partition function of QCD at non-zero baryon chemical 
potential is given by 
\be 
Z = \sum_i \exp\left(-\frac{E_i-\mu N_i}{T}\right),
\ee
where $i$ labels all quantum states of the system, $E_i$ and $N_i$ 
are the energy and baryon number of the state $i$. If the temperature 
and chemical potential are both zero then only the ground state 
contributes to the partition function. All other states give 
contributions that are exponentially small if the  volume of the 
system is taken to infinity. In QCD there is a massgap for states
that carry baryon number. As a a consequence there is an onset
chemical potential
\be 
\mu_{\it onset}=\min_i (E_i/N_i),
\ee 
such that the partition function is independent of $\mu$ for
$\mu<\mu_{\it onset}$. For $\mu>\mu_{\it onset}$ the baryon 
density is non-zero. If the chemical potential is just above 
the onset chemical potential we can describe QCD, to first 
approximation, as a dilute gas of non-interacting nucleons. 
In this approximation $\mu_{\it onset}=m_N$. Of course, the 
interaction between nucleons is essential. Without it, we 
would not have stable nuclei. As a consequence, nuclear matter
is self-bound and the energy per baryon in the ground state 
is given by
\be 
\frac{E_N}{N}-m_N \simeq -15\,{\rm MeV}.
\ee
The onset transition is a first order transition at which 
the baryon density jumps from zero to nuclear matter saturation
density, $\rho_0\simeq 0.14\,{\rm fm}^{-3}$. The first order 
transition continues into the finite temperature plane and 
ends at a critical endpoint at $T=T_c\simeq 10$ MeV, see
Fig.~\ref{fig_phase_1}. 

 Nuclear matter is a complicated many-body system and, 
unlike the situation at zero density and finite temperature, 
there is also no information from numerical simulations on 
the lattice. This is related to the so-called 'sign problem'.
At non-zero chemical potential the euclidean fermion determinant 
is complex and standard Monte-Carlo techniques based on importance
sampling fail. Recently, some progress has been made in 
simulating QCD for small $\mu$ and $T\simeq T_c$
\cite{Fodor:2001pe,deForcrand:2002ci,Allton:2002zi}, but the 
regime of small temperature remains inaccessible. As a consequence 
of the sign problem, there are also essentially no general 
results concerning the structure of the ground state. While 
the theorems of Vafa and Witten \cite{Vafa:tf,Vafa:1984xg}
rule out spontaneous breaking of parity or flavor at $T\neq 0$
and $\mu=0$, there are no theorems of this type at 
non-zero baryon density.

 However, if the density is very much larger than nuclear matter 
saturation density, $\rho\gg\rho_0$, we expect the problem to simplify. 
In this regime it is natural to use a system of non-interacting quarks 
as a starting point \cite{Collins:1974ky}. The low energy 
degrees of freedom are quark excitations and holes in the 
vicinity of the Fermi surface. Since the Fermi momentum is 
large, asymptotic freedom implies that the interaction between 
quasi-particles is weak. As a consequence, the naive expectation
is that chiral symmetry is restored and quarks and gluons are 
deconfined. It seems natural to assume that the quark liquid 
at high baryon density is continuously connected to the 
quark-gluon plasma at high temperature. These naive expectations
are summarized in the phase diagram shown in Fig.~\ref{fig_phase_1}. 

 Corrections to the non-interacting quark liquid can be 
studied in perturbation theory. The thermodynamic potential 
is given by \cite{Freedman:1976xs,Fraga:2001id}
\be
\label{omega_pqcd}
\Omega(\mu) = - \frac{N_f \mu^4}{4\pi^2}
\left\{1-2 \left(\frac{\alpha_s}{\pi}\right) 
- \left[G+N_f\log\left( \frac{\alpha_s}{\pi}\right)
 + \left(11-\frac{2}{3} N_f \right) 
   \log\left(\frac{\bar\Lambda}{\mu}\right) \right]
  \left(\frac{\alpha_s}{\pi}\right)^2 \right\} \; ,
\ee
where $G=G_0-0.536N_f+ N_f\ln{N_f}$, $G_0=10.374 \pm 0.13$.
Here, $\mu$ is the chemical potential for quark number. This 
convention is more natural in the context of perturbative 
QCD and we will use it for the remainder of these lectures.
Note that perturbative corrections reduce the pressure of 
the quark phase. At least qualitatively, this is agreement 
with the idea that at very low density the pressure of 
the hadron phase is bigger than the pressure of the quark
phase. 

\section{Color superconductivity}
\label{sec_csc}

  There are two problems with the perturbative expansion
equ.~(\ref{omega_pqcd}). One problem is related to the 
fact that while the electric gluon interaction is screened
by the mechanism discussed in Sect.~\ref{sec_screen} there 
is no screening of magnetic gluon exchanges. This not only 
implies that the magnetic sector of the theory becomes
non-perturbative, it also causes the Fermi liquid description 
to break down \cite{Holstein:1973,Reizer:1989}. The correction 
to the fermion self energy near the Fermi surface due to 
magnetic gluon exchanges is 
\cite{Baym:uj,Manuel:2000nh,Boyanovsky:2000bc,Brown:1999yd}
\be 
\label{nfl}
\Sigma_0(p_0) = \frac{g^2}{9\pi^2}
 \log\left(\frac{\mu}{p_0}\right).
\ee
This correction invalidates the Fermi liquid description
for energies $p_0\sim \mu \exp(-1/g^2)$. But even before
this phenomenon becomes important there is another 
effect that will invalidate the Fermi liquid picture. 
In Sect.~\ref{sec_bcs} we showed that the BCS instability
will lead to pair condensation whenever there is an 
attractive fermion-fermion interaction. At very large 
density, the attraction is provided by one-gluon exchange 
between quarks in a color anti-symmetric $\bar 3$ state. 
High density quark matter is therefore expected to behave 
as a color superconductor 
\cite{Frau_78,Barrois:1977xd,Bar_79,Bailin:1984bm}.

  Color superconductivity is described by a pair condensate
of the form
\be
\label{csc}
\Phi = \langle \psi^TC\Gamma_D\lambda_C\tau_F\psi\rangle.
\ee
Here, $C$ is the charge conjugation matrix, and $\Gamma_D,
\lambda_C,\tau_F$ are Dirac, color, and flavor matrices. 
Except in the case of only two colors, the order parameter
cannot be a color singlet. Color superconductivity is 
therefore characterized by the breakdown of color gauge 
invariance. This statement has to be interpreted in the 
sense of Sect.~\ref{sec_lg}. Gluons acquire a mass due
to the (Meissner-Anderson) Higgs mechanism. 

\begin{figure}
\includegraphics[width=12.0cm]{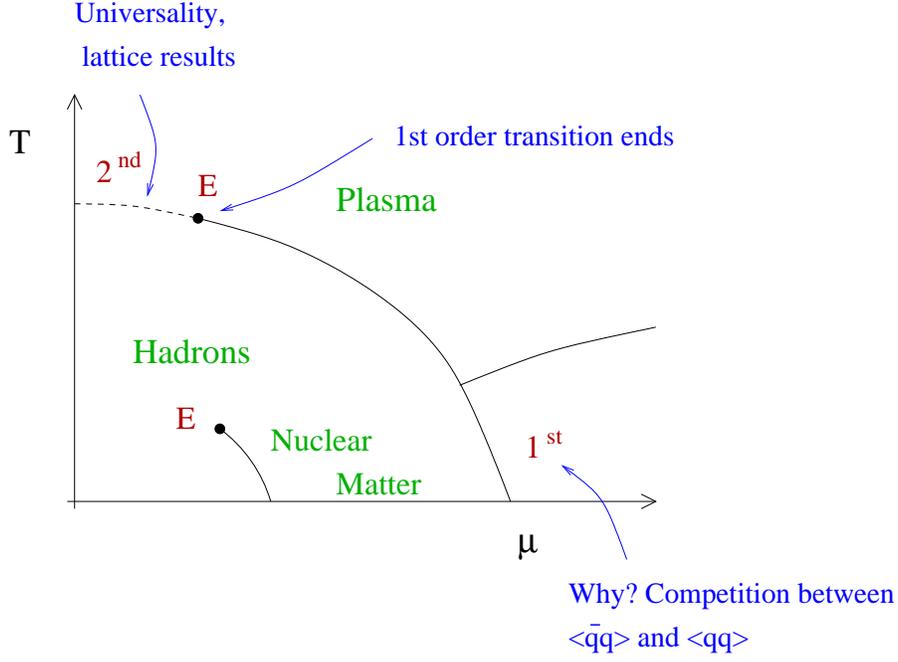}
\caption{\label{fig_phase_2}
First revision of the phase diagram of hadronic matter.
This figure shows the phase diagram of strongly interacting
matter obtained from a mean field treatment of chiral symmetry 
breaking and color superconductivity in QCD with two flavors, 
see e.g. \cite{Berges:1998}.}
\end{figure}

A rough estimate of the critical density for the transition
from chiral symmetry breaking to color superconductivity, the 
superconducting gap and the transition temperature is provided 
by schematic four-fermion models \cite{Alford:1998zt,Rapp:1998zu}.
Typical models are based on the instanton interaction 
\be 
\label{l_I}
{\cal L} = G_{I}\left\{
 (\bar\psi\tau^-_\alpha\psi)^2 + 
 (\bar\psi\gamma_5\tau^-_\alpha\psi)^2 
 \right\},
\ee
or a schematic one-gluon exchange interaction 
\be 
\label{l_OGE}
{\cal L} = G_{OGE}\left(\bar{\psi}\gamma_\mu
 \frac{\lambda^a}{2}\psi\right)^2 . 
\ee
Here $\tau^-_\alpha=(\vec{\tau},i)$ is an isospin matrix and
$\lambda^a$ are the color Gell-Mann matrices. The strength 
of the four-fermion interaction is typically tuned to reproduce 
the magnitude of the chiral condensate and the pion decay 
constant at zero temperature and density. In the mean field 
approximation the effective quark mass associated with chiral 
symmetry breaking is determined by a gap equation of the type 
\be 
\label{m_gap}
 M_Q = G_M  \int^\Lambda\frac{d^3p}{(2\pi)^3} 
  \frac{M_Q}{\sqrt{{\vec{p}}^{\,2}+M_Q^2}}
   \left(1-n_F(E_p)\right),
\ee
where $G_M$ is the effective coupling in the quark-anti-quark
channel and $\Lambda$ is a cutoff. Both the instanton interaction 
and the one-gluon exchange interaction are attractive in the color 
anti-triplet scalar diquark channel $\epsilon^{abc}(\psi^b C\gamma_5
\psi^c)$. A pure one-gluon exchange interaction leads to a 
degeneracy between scalar and pseudoscalar diquark condensation, 
but instantons are repulsive in the pseudoscalar diquark 
channel. The gap equation in the scalar diquark channel is 
\be
\label{d_gap}
 \Delta = \frac{G_D}{2} \int^\Lambda\frac{d^3p}{(2\pi)^3} 
  \frac{\Delta}{\sqrt{(|\vec{p}|-p_F)^2+\Delta^2}},
\ee
where we have neglected terms that do not have a singularity
on the Fermi surface, $|\vec{p}|=p_F$. In the case of a 
four-fermion interaction with the quantum numbers of 
one-gluon exchange $G_D=G_M/(N_c-1)$. The same result
holds for instanton effects. In order to determine 
the correct ground state we have to compare the 
condensation energy in the chiral symmetry broken 
and diquark condensed phases. We have ${\cal E} \sim 
f_\pi^2M_Q^2$ in the $(\bar{q}q)$ condensed phase 
and ${\cal E}\sim p_F^2\Delta^2/(2\pi^2)$ in the 
$(qq)$ condensed phase. 

 At zero temperature and density both equs.~(\ref{m_gap})
and (\ref{d_gap}) only have non-trivial solutions if 
the coupling exceeds a critical value. Since $G_M>G_D$ 
we have $M_Q>\Delta$ and the energetically preferred 
solution corresponds to chiral symmetry breaking. If 
the density increases Pauli-Blocking in equ.~(\ref{m_gap})
becomes important and the effective quark mass decreases. 
The diquark gap equation behaves very differently. 
Equ.~(\ref{d_gap}) has an infrared singularity on the 
Fermi surface, $p=p_F$, and this singularity is 
multiplied by a finite density of states, $N=p_F^2/(2\pi)^2$.
As a consequence, there is a non-trivial solution even 
if the coupling is weak. The gap grows with density
until the Fermi momentum becomes on the order of the 
cutoff. For realistic values of the parameters we find a first 
order transition for remarkably small values of the quark 
chemical potential, $\mu_Q\simeq 300$ MeV. The gap in 
the diquark condensed phase is $\Delta\sim 100$ MeV
and the critical temperature is $T_c\sim 50$ MeV.

 In the same model the finite temperature phase transition at 
zero baryon density is found to be of second order. This result 
is in agreement with universality arguments \cite{Pisarski:ms}
and lattice results. If the transition at finite density 
and zero temperature is indeed of first order then the 
first order transition at zero baryon density has to end in a 
tri-critical point 
\cite{Barducci:1989wi,Barducci:1989eu,Barducci:1993bh,Berges:1998,Halasz:1998}. 
The tri-critical point is quite remarkable, because it remains 
a true critical point, even if the quark masses are not zero. 
A non-zero quark mass turns the second order $T\neq 0$ transition 
into a smooth crossover, but the first order $\mu\neq 0$ transition 
persists. While it is hard to predict where exactly the tri-critical 
point is located in the phase diagram it may well be possible to 
settle the question experimentally. Heavy ion collisions at 
relativistic energies produce matter under the right conditions 
and experimental signatures of the tri-critical point have been 
suggested in \cite{Stephanov:1998}.

 A schematic phase diagram is shown in Fig.~\ref{fig_phase_2}. 
We should emphasize that this phase diagram is based on 
simplified models and that there is no proof that the 
transition from nuclear matter to quark matter along the 
$T=0$ line occurs via a single first order transition. 
Chiral symmetry breaking and color superconductivity 
represent two competing forms of order, and it seems 
unlikely that the two phases are separated by a second
order transition. However, since color superconductivity
modifies the spectrum near the Fermi surface, whereas 
chiral symmetry breaking operates near the surface of 
the Dirac sea, it is not clear that the two phases 
cannot coexist. Indeed, there are models in which 
a phase coexistence region appears \cite{Kitazawa:2002bc}.

\section{Phase structure in weak coupling}
\label{sec_phases}
\subsection{QCD with two flavors}
\label{sec_nf2}

  In this section we shall discuss how to use weak coupling
methods in order to explore the phases of dense quark matter.
We begin with what is usually considered to be the simplest 
case, quark matter with two degenerate flavors, up and down. 
Renormalization group arguments suggest 
\cite{Evans:1999ek,Schafer:1999na}, and explicit
calculations show \cite{Brown:1999yd,Schafer:2000tw}, that
whenever possible quark pairs condense in an $s$-wave. This
means that the spin wave function of the pair is anti-symmetric. 
Since the color wave function is also anti-symmetric, the Pauli
principle requires the flavor wave function to be anti-symmetric
too. This essentially determines the structure of the order
parameter \cite{Alford:1998zt,Rapp:1998zu}
\be
\label{2sc}
\Phi^a  = \langle \epsilon^{abc}\psi^b C\gamma_5 \tau_2\psi^c
 \rangle.
\ee
This order parameter breaks the color $SU(3)\to SU(2)$ and
leads to a gap for up and down quarks with two out of the 
three colors. Chiral and isospin symmetry remain unbroken. 

\begin{figure}
\includegraphics[width=11.0cm]{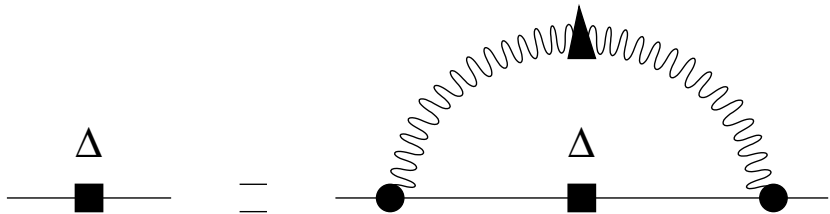}
\caption{\label{fig_gap}
Dyson-Schwinger (gap) equation in QCD at finite density. 
The square denotes an anomalous self energy (gap) insertion, 
and the triangle denotes a gluon self energy insertion. At 
leading order, quark self energy insertions or vertex 
corrections are not required \cite{Brown:1999yd}.}
\end{figure}

  We can calculate the magnitude of the gap and the 
condensation energy using weak coupling methods. In weak
coupling the gap is determined by ladder diagrams with 
the one gluon exchange interaction. These diagrams can 
be summed using the gap equation 
\cite{Son:1999uk,Schafer:1999jg,Pisarski:2000tv,Hong:2000fh,Brown:1999aq}
\bea
\label{eliash}
\Delta(p_4) &=& \frac{g^2}{12\pi^2} \int dq_4\int d\cos\theta\,
 \left(\frac{\frac{3}{2}-\frac{1}{2}\cos\theta}
            {1-\cos\theta+G/(2\mu^2)}\right. \\
 & & \hspace{3cm}\left.    +\frac{\frac{1}{2}+\frac{1}{2}\cos\theta}
            {1-\cos\theta+F/(2\mu^2)} \right)
 \frac{\Delta(q_4)}{\sqrt{q_4^2+\Delta(q_4)^2}}. \nonumber
\eea
Here, $\Delta(p_4)$ is the frequency dependent gap, $g$ is the 
QCD coupling constant and $G$ and $F$ are the self energies of
magnetic and electric gluons. This gap equation is very similar
to the BCS gap equation equ.~(\ref{d_gap}) obtained in four-fermion
models. The terms in the curly brackets arise from the magnetic
and electric components of the gluon propagator. The numerators
are the on-shell matrix elements ${\cal M}_{ii,00}=[\bar{u}_h(p_1)
\gamma_{i,0}u_h(p_3)][\bar{u}_h(p_2)\gamma_{i,0}u_h(p_4)]$ for 
the scattering of back-to-back fermions on the Fermi surface. 
The scattering angle is $\cos\theta=\vec{p}_1\cdot\vec{p}_3$. 
In the case of a spin zero order parameter, the helicity $h$ 
of all fermions is the same, see \cite{Schafer:1999jg} for 
more detail. 

 The main difference between equ.~(\ref{eliash}) and the 
BCS gap equation (\ref{d_gap}) is that because the gluon 
is massless, the gap equation contains a collinear $\cos\theta
\sim 1$ divergence. In a dense medium the collinear divergence is 
regularized by the gluon self energy. For $\vec{q}\to 0$ 
and to leading order in perturbation theory we have
\be
\label{pi_qcd}
 F = 2m^2, \hspace{1cm}
 G = \frac{\pi}{2}m^2\frac{q_4}{|\vec{q}|},
\ee
with $m^2=N_fg^2\mu^2/(4\pi^2)$. In the electric part,
$m_D^2=2m^2$ is the familiar Debye screening mass. In the 
magnetic part, there is no screening of static modes, 
but non-static modes are modes are dynamically screened
due to Landau damping. Equ.~(\ref{pi_qcd}) is, up to 
an overall degeneracy factor, exactly equal to the 
result obtained in Sect.~\ref{sec_screen}. The only 
difference is that in a relativistic theory the role 
of the tadpole graph in Fig.~\ref{fig_screen}b is played 
by the contribution of negative energy states in the 
particle-hole graph Fig.~\ref{fig_screen}a. We refer the 
reader to \cite{Blaizot:1993bb,Manuel:1995td,Rischke:2000qz}
for a more complete discussion of quasi-particle properties 
in a dense quark liquid. 

 For small energies dynamic screening of magnetic modes is
much weaker than Debye screening of electric modes. As a 
consequence, perturbative color superconductivity is dominated 
by magnetic gluon exchanges. Using equ.~(\ref{pi_qcd})
we can perform the angular integral in equ.~(\ref{eliash})
and find
\be
\label{eliash_mel}
\Delta(p_4) = \frac{g^2}{18\pi^2} \int dq_4
 \log\left(\frac{b\mu}{\sqrt{|p_4^2-q_4^2|}}\right)
    \frac{\Delta(q_4)}{\sqrt{q_4^2+\Delta(q_4)^2}},
\ee
with $b=256\pi^4(2/N_f)^{5/2}g^{-5}$. We can now see why 
it was important to keep the frequency dependence of the 
gap. Because the collinear divergence is regulated by
dynamic screening, the gap equation depends on $p_4$
even if the frequency is small. We can also see that
the gap scales as $\exp(-c/g)$. The collinear divergence 
leads to a gap equation with a double-log behavior. 
Qualitatively
\be
\label{dlog}
 1 \sim \frac{g^2}{18\pi^2}
 \left[\log\left(\frac{\mu}{\Delta}\right)\right]^2,
\ee
from which we conclude that $\Delta\sim\exp(-c/g)$. 
The approximation equ.~(\ref{dlog}) is not sufficiently
accurate to determine the correct value of the 
constant $c$. A more detailed analysis shows that
the gap on the Fermi surface is given by
\be
\label{gap_oge}
\Delta_0 \simeq 512\pi^4(2/N_f)^{5/2}b_0'\mu g^{-5}
   \exp\left(-\frac{3\pi^2}{\sqrt{2}g}\right).
\ee
The factor $b_0'$ is related to non-Fermi liquid
effects, see equ.~(\ref{nfl}). Note that since $\Delta
\sim \exp(-1/g)$ non-Fermi liquid effects are indeed 
sub-leading. In perturbation theory $b_0'=\exp(-(\pi^2+4)
(N_c-1)/16)$ \cite{Brown:1999aq,Wang:2001aq}. The 
condensation energy is given by
\be
\epsilon = -N_d \Delta_0^2\left(\frac{\mu^2}{4\pi^2}\right),
\ee
where $N_d=4$ is the number of condensed species. The
critical temperature is $T_c/\Delta_0 =e^\gamma/\pi\simeq
0.56$, as in standard BCS theory. For chemical 
potentials $\mu<1$ GeV, the coupling constant is not small 
and the applicability of perturbation theory is in doubt. 
If we ignore this problem and extrapolate the perturbative 
calculation to densities $\rho\simeq 5\rho_0$ we find gaps 
$\Delta\simeq 100$ MeV. This result is in surprisingly good 
agreement with the estimates from Nambu-Jona-Lasinio models 
discussed in Sect.~\ref{sec_csc}. 

 We note that the 2SC phase defined by equ.~(\ref{2sc})
has two gapless fermions and an unbroken $SU(2)$ gauge 
group. The gapless fermions are singlets under the 
unbroken $SU(2)$. As a consequence, we expect the 
$SU(2)$ gauge group to become non-perturbative. An 
estimate of the $SU(2)$ confinement scale was given 
in \cite{Rischke:2000cn}. We also note that even though
the Copper pairs carry electric charge the $U(1)$
of electromagnetism is not broken. The generator of
this symmetry is a linear combination of the original 
electric charge operator and the diagonal color charges. 
Under this symmetry the gapless fermions carry the 
charges of the proton and neutron. Possible pairing between 
the gapless fermions was discussed in 
\cite{Alford:1998zt,Alford:2002xx}.
 
\begin{figure}
\includegraphics[width=11.0cm]{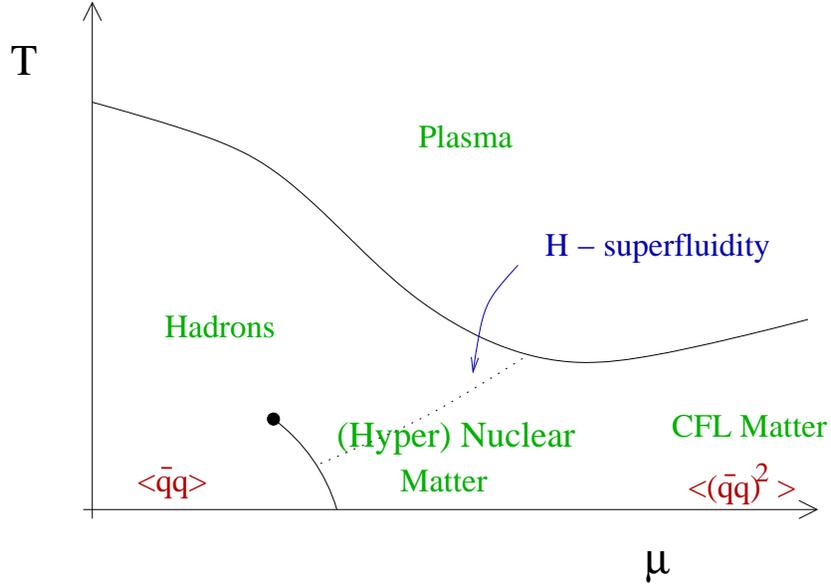}
\caption{\label{fig_phase_3}
Conjectured phase diagram of $N_f=3$ hadronic matter in the 
limit of exact flavor symmetry.}
\end{figure}

\subsection{QCD with three flavors: Color-Flavor-Locking}
\label{sec_cfl} 

 If quark matter is formed at densities several times 
nuclear matter density we expect the quark chemical 
potential to be larger than the strange quark mass.
We therefore have to determine the structure of the 
superfluid order parameter for three quark flavors.
We begin with the idealized situation of three 
degenerate flavors. From the arguments given in the 
last section we expect the order parameter to be 
color and flavor anti-symmetric matrix of the form
\be
\label{order}
  \Phi^{ab}_{ij}=
  \langle \psi^a_i C\gamma_5\psi^b_j\rangle.
\ee
In order to determine the precise structure of this
matrix we have to extremize grand canonical potential.
We find \cite{Schafer:1999fe,Evans:1999at}
\be
\label{cfl}
\Delta^{ab}_{ij} = 
 \Delta_A (\delta_i^a\delta_j^b-\delta_i^b\delta_j^a)
+\Delta_S (\delta_i^a\delta_j^b+\delta_i^b\delta_j^a),
\ee
which describes the color-flavor locked (CFL) phase proposed 
in \cite{Alford:1999mk}. In the weak coupling limit $\Delta_S 
\ll\Delta_A$ and $\Delta_A=2^{-1/3}\Delta_0$ where $\Delta_0$ 
is the gap in the 2SC phase, equ.~(\ref{gap_oge}) 
\cite{Schafer:1999fe}. In the CFL phase both color and flavor 
symmetry are completely broken. There are eight combinations 
of color and flavor symmetries that generate unbroken global 
symmetries. The unbroken symmetries are
\be
\psi^a_{L,i}\to (U^*)^{ab}U_{ij}\psi^b_{Lj},
\hspace{1cm}
\psi^a_{R,i}\to (U^*)^{ab}U_{ij}\psi^b_{Rj},
\ee
for $U\in SU(3)_V$. The symmetry breaking pattern is 
\be
\label{sym_3}
SU(3)_L\times SU(3)_R\times U(1)_V\to SU(3)_V .
\ee
We observe that color-flavor-locking implies that chiral
symmetry is broken. The mechanism for chiral symmetry 
breaking is quite unusual. The primary order parameter 
$\langle \psi^a_{Li}C\Delta^{ab}_{ij}\psi^b_{Lj}\rangle
=-\langle \psi^a_{Ri}C\Delta^{ab}_{ij}\psi^b_{Rj}\rangle$
involves no coupling between left and right handed fermions.
In the CFL phase both left and right handed flavor are locked 
to color, and because of the vectorial coupling of the gluon 
left handed flavor is effectively locked to right handed flavor. 
Chiral symmetry breaking also implies that $\langle \bar{\psi}
\psi\rangle$ has a non-zero expectation value. We shall compute 
the quark condensate in Sect.~\ref{sec_inst}. In the CFL phase 
$\langle \bar{\psi}\psi\rangle^2\ll \langle(\bar{\psi}\psi)^2
\rangle$. Another measure of chiral symmetry breaking is 
provided by the pion decay constant. In Sect.~\ref{sec_hdet}
we will show that in the weak coupling limit $f_\pi^2$ is 
proportional to the density of states on the Fermi surface.

 The symmetry breaking pattern $SU(3)_L\times SU(3)_R \to SU(3)_V$ 
in the CFL phase is identical to the symmetry breaking pattern
in QCD at low density. The spectrum of excitations in the 
color-flavor-locked (CFL) phase also looks remarkably like the 
spectrum of QCD at low density \cite{Schafer:1999ef}. The 
excitations can be classified according to their quantum 
numbers under the unbroken $SU(3)$, and by their electric 
charge. The modified charge operator that generates a true 
symmetry of the CFL phase is given by a linear combination 
of the original charge operator $Q_{em}$ and the color hypercharge 
operator $Q={\rm diag}(-2/3,-2/3,1/3)$. Also, baryon number is only 
broken modulo 2/3, which means that one can still distinguish baryons 
from mesons. We find that the CFL phase contains an octet of Goldstone 
bosons associated with chiral symmetry breaking, an octet of vector 
mesons, an octet and a singlet of baryons, and a singlet Goldstone 
boson related to superfluidity. All of these states have integer 
charges.  

\begin{figure}
\includegraphics[width=7.5cm]{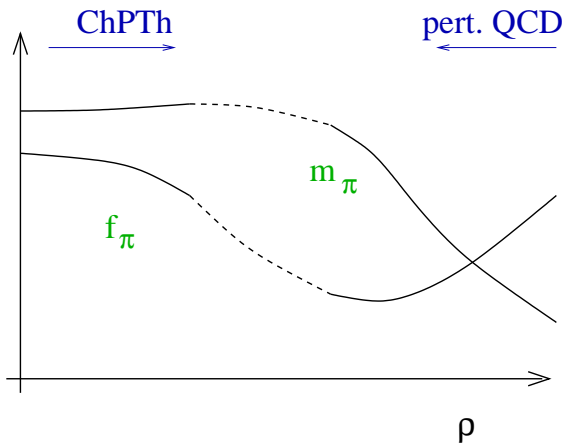}
\caption{\label{fig_had_par}
Schematic plot of the properties of hadronic modes in $N_f=3$ 
QCD in the limit of exact flavor symmetries. Hadronic parameters 
can be determined using chiral perturbation theory at low density, 
and using perturbative QCD at high density. The high density 
behavior is studied in Sect.~\ref{sec_hdet}.}
\end{figure}

  With the exception of the $U(1)$ Goldstone boson, these states
exactly match the quantum numbers of the lowest lying multiplets
in QCD at low density. In addition to that, the presence of the 
$U(1)$ Goldstone boson can also be understood. The $U(1)$ order
parameter is $\langle (uds)(uds)\rangle$. This order parameter
has the quantum numbers of a $0^+$ $\Lambda\Lambda$ pair condensate.
In $N_f=3$ QCD, this is the most symmetric two nucleon channel, 
and a very likely candidate for superfluidity in nuclear matter
at low to moderate density. We conclude that in QCD with three
degenerate light flavors, there is no fundamental difference 
between the high and low density phases. This implies that 
a low density hyper-nuclear phase and the high density quark 
phase might be continuously connected, without an intervening 
phase transition. A conjectured phase diagram is shown in 
Fig.~\ref{fig_phase_3}. 

 An important consistency check for the structure of the phase 
diagram is provided by anomaly matching arguments. Anomaly matching 
expresses the requirement that the flavor anomalies of the microscopic 
theory can be represented in the effective theory for the low energy 
degrees of freedom \cite{tHooft:1980,Peskin:1982}. It was shown in 
\cite{Sannino:2000kg,Sannino:2003pq} that this requirement
also applies to gauge theories at finite baryon density. We 
observe that color-flavor-locking realizes the standard 
Goldstone boson option for anomaly matching in $N_f=3$ QCD,
whereas the 2SC phase corresponds to the massless proton 
and neutron option in $N_f=2$ QCD.

 We also note that the CFL phase provides 
a weak coupling realization of a phase with chiral symmetry breaking 
and a mass gap. This means that the CFL phase offers the opportunity
to study many of the 'hard' problems of non-perturbative QCD
in a perturbative setting. For example, we can compute hadronic
parameters at high baryon density and try to extrapolate the results 
to low density, see Fig.~\ref{fig_had_par}.

\subsection{$N_f\neq 2,3$}
\label{sec_csl}
 
 Color-flavor locking can be generalized to QCD with more than
three flavors \cite{Schafer:1999fe}. For all $N_f\geq 3$ the 
high density phase is fully gapped. Also, at least part of the 
chiral $SU(N_f)\times SU(N_f)$ symmetry is broken for all $N_f\ge 3$, 
but only in the case $N_f=3$ do we find the $T=\mu=0$ pattern of 
chiral symmetry breaking, $SU(N_f)_L \times SU(N_f)_R\to SU(N_f)_V$. 

  While the case $N_f>3$ is mostly of academic interest, the 
phase structure of $N_f=1$ QCD is possibly relevant to real 
QCD. Because of mass effects or non-zero electron chemical 
potentials the Fermi surface of strange and non-strange quarks 
may get pushed too far apart for strange-non-strange pairing
to occur, see Sect.~\ref{sec_unlock}. As a consequence, there
may be regions in the phase diagram where $(ss)$ pairing occurs.
In the case of single flavor pairing the order parameter is 
flavor-symmetric and the Cooper pairs carry non-zero angular 
momentum. The simplest order parameters are of the form
\be
\label{spin1}
\vec\Phi_1^a=
   \langle \epsilon^{abc}\psi^b C\vec\gamma \psi^c\rangle,
\hspace{1cm}
\vec\Phi_2^a=
   \langle \epsilon^{abc}\psi^b C\hat{q} \psi^c\rangle.
\ee
The corresponding gaps can be determined using the methods
introduced in section \ref{sec_nf2}. We find $\Delta(\Phi_{1,2})
=\exp(-3c_{1,2})\Delta_0$ where $c_1=-1.5$, $c_2=-2$ and $\Delta_0$ 
is the spin zero gap \cite{Brown:1999yd,Schafer:2000tw}. While the 
natural scale of the s-wave gap is $\Delta_0\simeq 100$ MeV, the 
p-wave gap is expected to be less than 1 MeV.

 The spin one order parameter equ.~(\ref{spin1}) is a color-spin
matrix. This opens the possibility that color and spin degrees
become entangled, similar to the color-flavor-locked phase
or the B-phase of liquid $^3$He. The corresponding order 
parameter is
\be
\label{csl}
 \Phi_{CSL} = \delta^a_i \langle\epsilon^{abc}\psi^b C
 \left(\cos(\beta)\hat{q}_i+\sin(\beta)\gamma_i \right)
 \psi^c\rangle,
\ee
where the angle $\beta$ determines the mixing between the
two types of condensates shown in equ.~(\ref{spin1}). A
weak coupling analysis of the effective potential shows that 
the color-spin-locked phase equ.~(\ref{csl}) is favored over 
the ``polar'' phase equ.~(\ref{spin1}) \cite{Schafer:2000tw}.
The value of $\beta$ depends sensitively on the interaction 
and the mass of the quark. In the color-spin-locked phase 
color and rotational invariance are broken, but a diagonal 
$SO(3)$ survives. As a consequence, the gap is isotropic. 
Color-spin-locking also leads to an unusual spectrum of 
quasi-particles. In the Fermi liquid phase there is a left 
and right handed color triplet of quarks. In the CSL phase 
we find a spin 3/2 quartet and a spin 1/2 doublet of the 
unbroken $SO(3)$ symmetry. The CSL phase 'knows' that 
one-flavor nuclear matter consists of spin 3/2 delta baryons.

\subsection{$N_c= 2$}
\label{sec_nc2}
 
 QCD with $N_c=2$ colors is an interesting model system. The 
interest in this theory derives from the fact that the determinant
of the euclidean Dirac operator in QCD with $N_c=2$ remains real
even if the baryon chemical potential is non-zero. This means
that two-color QCD with an even number of flavors can be studied 
on the lattice using standard techniques. In addition to that, 
there is theoretical control not only in the regime of large density,
but also in the regime of small density. 

 For simplicity we will concentrate on the case of $N_f=2$ 
flavors. $SU(2)$ gauge theory has a meson spectrum which is 
very similar to three-color QCD. Baryons, on the other hand, are 
bosons rather than fermions and their spectrum is very different 
as compared to $N_c=3$ QCD. Two-color QCD is also characterized
by an enlarged chiral symmetry. We can write 
\be
\Psi = \left(\begin{array}{c} 
  \psi_L \\ \sigma_2\tau_2 \psi_R^* 
  \end{array} \right),
\ee
where $\sigma_2,\tau_2$ are anti-symmetric color and flavor 
$SU(2)$ matrices. Two-color QCD is not only invariant under 
$SU(2)_L\times SU(2)_R$ transformations acting on the upper
and lower components of $\Psi$ separately, but under the full 
$SU(4)$ group \cite{Peskin:1980gc,Rapp:1998zu,Kogut:1999iv}. The 
$SU(4)$ chiral symmetry mixes the quark-anti-quark condensate 
$\langle\bar{\psi}^a\psi^a\rangle$ with the diquark condensate 
$\langle\epsilon^{ab}\psi^{a\,T}C\gamma_5\tau_2 \psi^b\rangle$. 

 At zero temperature and density, and in the presence of a 
small quark mass, the chiral $SU(4)$ symmetry is broken to
$Sp(4)$ by a quark-anti-quark condensate $\langle\bar{\psi}^a\psi^a
\rangle$. There are 5 Goldstone bosons, three pions $\vec{\pi}$,
the scalar diquark $S$ and the scalar anti-diquark $\bar{S}$.
If we turn on a baryon chemical potential then the scalar 
diquark will Bose condense if the chemical potential exceeds
the mass of the diquark. Since the scalar diquark is a Goldstone
boson, this phenomenon can be studied using the chiral effective 
lagrangian of $N_c=2$ QCD. The effective lagrangian is given
by \cite{Kogut:1999iv,Kogut:2000ek}
\be
\label{l_nc2}
{\cal L} = \frac{f_\pi^2}{4} {\rm Tr}\left[
 D_\mu\Sigma D^\mu\Sigma^\dagger\right] 
 +\Big[ mC {\rm Tr}(\hat{M}\Sigma^\dagger) + h.c. \Big]
+ \ldots.
\ee
The chiral field $\Sigma$ parametrizes the coset $SU(4)/Sp(4)$.
$\Sigma$ is an anti-symmetric unitary matrix. It transforms as 
$\Sigma\to U\Sigma U^T$ under the chiral $SU(4)$ symmetry. The 
covariant derivative is defined as
\be 
iD_\nu  = i\partial_\nu -\mu\delta_{\nu 0}
 \left(\hat{B}\Sigma+\Sigma\hat{B}^T\right).
\ee
The matrices $\hat{M}$ and $\hat{B}$ are determined by the 
transformation properties of the mass term $m\bar{\psi}\psi$
and the chemical potential term $\mu\psi^\dagger\psi$ under
the chiral $SU(4)$ symmetry. We have
\be
\hat{M} =\left( \begin{array}{cc} 
0 & 1 \\ -1 & 0
\end{array}\right), \hspace{1cm}
\hat{B} =\left( \begin{array}{cc} 
1 & 0 \\ 0 & -1
\end{array}\right).
\ee
We can determine the ground state by minimizing the potential
\be
\label{veff_nc2}
V(\Sigma) = -\frac{f_\pi^2}{2} {\rm Tr}\left[
 \Sigma B^T \Sigma^\dagger B\right] 
 +\Big[ mC {\rm Tr}(\hat{M}\Sigma^\dagger) + h.c. \Big].
\ee
For zero density and finite quark mass the minimum is 
$\Sigma_{\bar{q}q}=\hat{M}$. In this state $SU(4)$ is 
broken to $Sp(4)$. Goldstone bosons are described by 
fluctuations $\Sigma = U\Sigma_{\bar{q}q} U^T$ with 
$U=\exp(i\phi^a X^a/f_\pi)\in SU(4)/Sp(4)$. Here, $X^a$ 
are the $SU(4)$ generators that act non-trivially on 
$\Sigma_{\bar{q}q}$. As mentioned above, there a 5 
Goldstone modes. 

 The first term in equ.~(\ref{veff_nc2}) is minimized by
\be
\Sigma_{qq} = \left( \begin{array}{cc} 
\sigma_2 & 0\\ 0 & \sigma_2
\end{array}\right).
\ee
For non-zero chemical we expect the minimum to be of the 
form $\Sigma_0=\Sigma_{\bar{q}q}\cos(\alpha)+\Sigma_{qq}
\sin(\alpha)$. Substituting this ansatz into equ.~(\ref{veff_nc2})
we find $\alpha=0$ for $\mu<m_\pi/2$ and $\cos(\alpha)=m_\pi^2
/(4\mu)^2$. Differentiating the effective potential with 
respect to the chemical potential we find the baryon density
\be
\label{rho_nc2}
\rho_B = 8f_\pi^2\mu\left(1-\left(\frac{m_\pi^2}{4\mu^2}
 \right)^2\right).
\ee
For small $\mu=m_\pi/2+\delta\mu$ this result is exactly 
of the same form as equ.~(\ref{rho_bose}). This implies that
the physical phenomenon is indeed Bose condensation of 
scalar diquarks interacting through a short range repulsive
interaction. 

 If the density becomes large, $\mu\sim m_\rho\gg m_\pi$, the
effective lagrangian description breaks down. On the other 
hand, if $\mu\gg\Lambda_{QCD}$ we expect to find a perturbative
BCS superfluid of diquark pairs. The gap and the superfluid
condensate can be computed using the methods discussed 
in Sect.~\ref{sec_nf2}. We find 
\be
\label{gap_nc2}
 \Delta = 512\pi^4 b_0'\mu g^{-5}
 \exp\left(-\frac{2\pi^2}{g}\right),
\ee
with $b_0'=\exp(-(\pi^2+4)/16)$. Numerical studies 
on the lattice can be used in order to verify the limiting 
behavior at small and large density, and to study the 
Bose condensation/BCS crossover, see \cite{Kogut:2002cm}
and references therein. Similar studies have also been 
performed in the instanton liquid model \cite{Schafer:1998up}.
There are many other interesting questions that can be studied 
in two-color QCD. It was suggested, for example, that 
vector mesons (diquarks) may condense if the chemical 
potential is on the order of the vector meson mass
\cite{Lenaghan:2001sd,Sannino:2002wp}. Other interesting
questions concern the structure of the phase diagram at 
non-zero temperature \cite{Splittorff:2002xn}, and the nature 
of the deconfinement transition. It was also pointed out that 
the behavior of the $\eta'$ mass in two-color QCD can be used 
in order to study the mechanism of $U(1)_A$ breaking in QCD 
\cite{Schafer:2002yy}. Finally, we should mention that there
are some other gauge theories in which the euclidean fermion
determinant is positive even if the chemical potential is 
non-zero. These theories include QCD at finite isospin chemical 
potential and QCD with a non-zero density of quarks in the adjoint 
representation of color \cite{Alford:1998sd,Son:2000xc,Kogut:2000ek}.
It is amusing that all of these theories have Goldstone bosons
that carry the conserved charge. As a consequence, the low 
density state of these theories is a dilute Goldstone boson
condensate, similar to the $N_c=2$ diquark condensate studied
in this section.

\subsection{$N_c\to\infty$}
\label{sec_nc}
 
 In the large $N_c$ limit quark-quark scattering is 
suppressed as compared to quark-quark-hole scattering. 
The color factors in the two channels are
\be
c=\frac{N_c+1}{2N_c}   \hspace{0.5cm} (qq),\hspace{1cm}
c=\frac{N_c^2-1}{2N_c} \hspace{0.5cm} (qh),
\ee
suggesting that particle-particle pairing, and superconductivity,
is disfavored at large $N_c$. As a consequence, other forms of
pairing may take place. Particle-hole scattering is not 
suppressed, but particle-hole pairing can only take place
over a small part of the Fermi surface. The order parameter
for particle-hole pairing is \cite{Deryagin:1992}
\be 
\langle\bar\psi(x)\psi(y) \rangle = 
 \exp(i\vec{p}\cdot(\vec{x}+\vec{y}))\Sigma(x-y),
\ee
where $|\vec{p}|=p_F$ is a vector on the Fermi surface. This 
state describes a chiral density wave. It resembles a charge
or spin density wave in quasi-one-dimensional condensed
matter systems \cite{Gruner}. QCD, of course, is not 
quasi-one-dimensional but at large $N_c$ screening due 
to fermions is weak and the perturbative one-gluon exchange 
interaction is very strongly dominated by collinear 
scattering. The problem was studied in the weak coupling
approximation in \cite{Shuster:1999}. It was found that 
the transition from color superconductivity to chiral 
density waves requires very large values of $N_c>1000$.
On the other hand, if the coupling is strong and the 
density is not too large, then the chiral density wave
state may compete with color superconductivity even for
three colors \cite{Rapp:2000zd}. 

\section{The role of the strange quark mass}
\label{sec_ms} 
\subsection{BCS theory: toy model}
\label{sec_unlock}
 
    At baryon densities relevant to astrophysical objects dis\-tor\-tions 
of the pure CFL state due to non-zero quark masses cannot be neglected 
\cite{Alford:1999pa,Schafer:1999pb,Rapp:1999qa,Bedaque:1999nu,Alford:2000ze,Schafer:2000ew,Bedaque:2001je,Kaplan:2001qk,Casalbuoni:2002st,Fugleberg:2002rk,Casalbuoni:2003cs}.
The most important effect of a non-zero strange quark mass is that 
the light and strange Fermi surfaces will no longer be of equal size. 
When the mismatch is much smaller than the gap one calculates
assuming degenerate quarks, we might expect that it has very little 
consequence, since at this level the original particle and hole 
states near the Fermi surface are mixed up anyway. On the other 
hand, when the mismatch is much larger than the nominal gap,
we might expect that the ordering one would obtain for degenerate
quarks is disrupted, and that to a first approximation
one can treat the light and heavy quark dynamics separately.

\begin{figure}
\includegraphics[width=13.0cm]{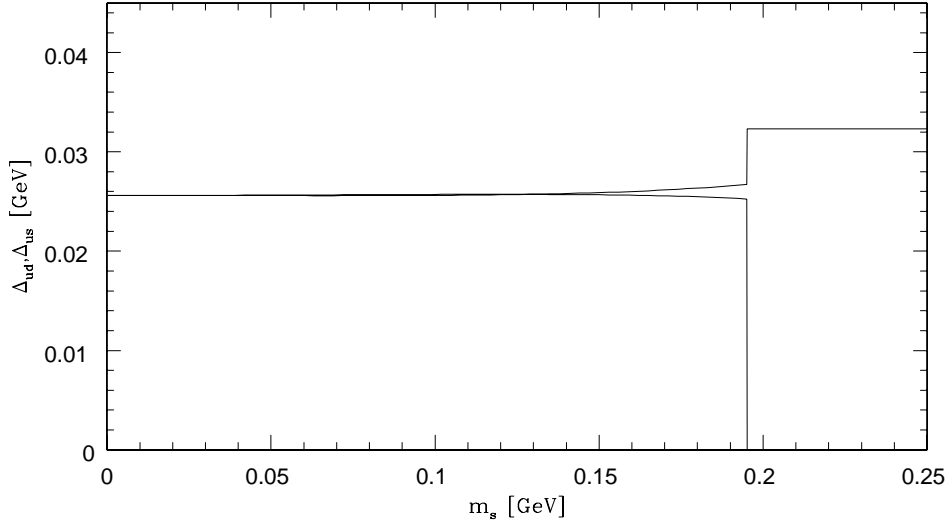}
\caption{\label{fig_unlock}
Light quark gap $\Delta_{ud}$ and strange-non-strange 
gap $\Delta_{us}$ as a function of the strange quark mass
for $\mu=0.5$ GeV. This figure shows the result of a mean 
field analysis based on a schematic BCS interaction 
\cite{Schafer:1999pb}. The interaction strength was
(arbitrarily) adjusted to $\Delta_{ud}(m_s\!=\!0)=25$
MeV. }
\end{figure}

 We can see this in a more quantitative fashion by studying a
schematic gap equation that describes the spin singlet pairing of
two fermions with different masses. In a basis of particles of
the first kind and holes of the second the quadratic part of the
action is
\be
\label{gorkov}
{\cal S} = \int\frac{d^4p}{(2\pi)^4}
 \left(\psi^\dagger_{(1)}\hspace{0.3cm} \psi_{(2)}\right)
 \left(\begin{array}{cc}
     p_0-\epsilon_p^1  & \Delta \\
     \Delta^* & p_0+\epsilon_p^2
 \end{array}\right)
 \left( \begin{array}{c}
      \psi_{(1)} \\ \psi^\dagger_{(2)}
 \end{array}\right) .
\ee
Here, $\epsilon^{1,2}_p=E^{1,2}_p-\mu$ and $E^{1,2}_p=(p^2+
m_{1,2}^2)^{1/2}$ where $m_{1,2}$ are the masses of particle 
one and two. The particle and hole propagators are determined 
by the inverse of the matrix equ.~(\ref{gorkov}). The off-diagonal 
(anomalous) propagator is
\be
G_{21}=\frac{\Delta}{(p_0-\epsilon^1_p)(p_0+\epsilon^2_p)-\Delta^2}.
\ee
We study the effect of a zero range interaction $G(\psi_1
\sigma_2\psi_2)(\psi^\dagger_1\sigma_2\psi^\dagger_2)$. The
pairing is described by the gap equation
\be
 \Delta= G\int\frac{d^4p}{(2\pi)^4}\frac{\Delta}
 {(p_0+R+i\delta {\rm sgn}(p_0))^2-\bar\epsilon_p^2-\Delta^2}.
\ee
Here, we have introduced $\bar\epsilon_p=\bar E_p-\mu=(\epsilon^1_p
+\epsilon^2_p)/2$ and $R=(\epsilon^1_p-\epsilon^2_p)/2$. In practice,
we are interested in pairing between almost massless up or down quarks
and massive strange quarks. In that case, $R\simeq m_s^2/(4p_F)\simeq
m_s^2/(4\mu)$. The poles of the anomalous propagator are located at
$p_0=-R\pm(\bar\epsilon^2_p+\Delta^2)^{1/2}-i {\rm sgn}(p_0)$. As usual, 
we close the integration contour in the lower half plane. Let us denote
the solution of the gap equation in the case $R=0$ by $\Delta_0$.
Then, if $R<\Delta_0$, the pole with the positive sign of the square
root is always included in the integration contour and we have
\be
\label{gap}
 \Delta = \frac{G\mu^2}{4\pi^2} \int d\bar\epsilon_p\,
 \frac{\Delta}{\sqrt{\bar\epsilon^2_p+\Delta^2}}.
\ee
This result is, up to a small correction in the density of states
that we have neglected here, identical to the gap equation for
degenerate fermions, so $\Delta \approx \Delta_0$. If, on the other hand,
$R>\Delta_0$ there only is a pole in the lower half plane if
$\bar\epsilon_p>\sqrt{R^2-\Delta^2}$. Carrying out the $p_0$
integration again leads to the gap equation (\ref{gap}), but
with the $\bar\epsilon_p$ integration restricted by the condition
just mentioned. This cuts out the infrared singularity at
$\bar\epsilon_p=0$ and one can easily verify that the gap equation
does not have a non-trivial solution for weak coupling. We 
thus conclude that a necessary condition for pairing is that 
\be
 m_s^2<m_s^2({\it crit}) \simeq 4p_F\Delta(m_s\!=\!0).
\ee

\subsection{BCS theory: CFL phase}
\label{sec_cfl_ms}
 
So far, we have only dealt with a simple pair condensate involving
strange and non-strange quarks. In practice, we are interested in a
somewhat more complicated situation. In particular, we want to consider
the transition between the color-flavor locked phase for small $m_s$
and the two-flavor color superconductor in the limit of large $m_s$.
This analysis can be carried out along the same lines as the toy
model discussed above. We now consider the following free action
\cite{Schafer:1999pb,Alford:1999pa}
\be
S = \int \frac{d^4p}{(2\pi)^4}
  \left(\psi^\dagger\hspace{0.3cm} \psi\right)
\left(\begin{array}{cc}
  (p_0-\epsilon_p) X_1 - 2RX_s & \Delta_{ud}X_{ud}+\Delta_{us}X_{us} \\
  \Delta_{ud}X_{ud}+\Delta_{us}X_{us} & (p_0+\epsilon_p) X_1+2RX_s
\end{array}\right)
\left( \begin{array}{c}
  \psi  \\ \psi^\dagger
\end{array}\right) ,
\label{gor_cfl}
\ee
where $\psi$ is now a 9 component color-flavor spinor.
$X_1, X_s, X_{ud}$ and $X_{us}$ are color-flavor matrices
\be
\begin{array}{rclrcl}
X_1    &=& \delta^{\alpha\beta}\delta_{ab}, \hspace{1cm}& 
X_s    &=& \delta^{\alpha\beta}\delta_{a3}\delta_{b3}\\
X_{ud} &=& \epsilon^{3\alpha\beta}\epsilon_{3ab}, &
X_{us} &=& \epsilon^{2\alpha\beta}\epsilon_{2ab}
          +\epsilon^{1\alpha\beta}\epsilon_{1ab},
\end{array}
\ee
where $\alpha,\beta$ are color, and $a,b$ flavor indices.
$\Delta_{ud}$ is the gap for $\langle ud\rangle$ condensation, and
$\Delta_{us}$ is the gap for $\langle us\rangle=\langle ds\rangle$
condensation. Color-flavor locking corresponds to the case $\Delta_{ud}
=\Delta_{us}$, and the two flavor superconductor corresponds to
$\Delta_{us}=0,\,\Delta_{ud}\neq 0$. 

  Flavor symmetry breaking is again caused by $R\simeq m_s^2/(4p_F)$. 
The Nambu-Gorkov matrix equ.~(\ref{gor_cfl}) can be diagonalized 
exactly. The eigenvalues and their degeneracies are
\be
\begin{array}{ll}
p_0 \pm  \left(\epsilon_p^2+\Delta_{ud}^2\right)^{1/2},  & d=3 \\
p_0 - R \pm \left(\bar\epsilon_p^2+\Delta_{us}^2\right)^{1/2}, & d=2\\
p_0 + R \pm \left(\bar\epsilon_p^2+\Delta_{us}^2\right)^{1/2}, & d=2\\
p_0 \pm \left( \epsilon_p^2 + 2R\epsilon_p +2R^2 + 2\Delta_{us}^2
   +\frac{1}{2}\Delta_{ud}^2 \pm\frac{1}{2}S\right)^{1/2},     & d=1,1\\
\end{array}
\label{evals}
\ee
where
\be
S=\left(8\Delta_{us}^2\left(\Delta_{ud}^2+4R^2\right)
+\left(\Delta_{ud}^2-4R(\epsilon_p+R)\right)^2\right)^{1/2}.
\ee
The result becomes easier to understand if we consider some simple 
limits. If we ignore flavor symmetry breaking, $R=0$, and set 
$\Delta_{ud}=\Delta_{us}$ we find 8 eigenvalues $p_0\pm(\epsilon_p^2
+\Delta^2)^{1/2}$ and one eigenvalue with the gap $2\Delta$. These
states fill out an octet and a singlet of the unbroken $SU(3)_F$ 
symmetry of the CFL phase. If, on the other hand, we set $\Delta_{us}
=0$ we find 4 eigenvalues $p_0\pm(\epsilon_p^2+\Delta^2)^{1/2}$ 
while the other 5 eigenvalues have vanishing gaps. This is the 
spectrum of the $N_f=2$ phase. If flavor symmetry is broken, 
we find that the $SU(3)$ octet splits into two $SU(2)$ doublets, 
one triplet and singlet. 

  We note that in the presence of flavor symmetry breaking the first
three eigenvalues, which depend on $\Delta_{ud}$ only, are completely
unaffected. For the next 4 eigenvalues, which only depend on
$\Delta_{us}$, the energy $p_0$ is effectively shifted by $R$. This
is exactly as in the simple toy model discussed above. It implies that
for $R>\Delta_{us}$, when we close the integration contour in the
complex $p_0$ plane, we do not pick up this pole. The last two
eigenvalues are more complicated. They depend on both $\Delta_{ud}$
and $\Delta_{us}$, and they explicitly contain the flavor symmetry
breaking parameter $R$. Nevertheless, the structure of the eigenvalues
is certainly suggestive of the idea that for $R<\Delta_{us}^0$ we
have $\Delta_{us}\simeq \Delta_{ud}$, and the gaps are almost
independent of $R$, while at $R\simeq\Delta_{us}^0$ there is a
discontinuity and $\Delta_{us}$ goes to zero.

This is borne out by a more detailed calculation. For this purpose,
we add a flavor and color anti-symmetric short range interaction
\be
{\cal L} = \frac{K}{4}\big( \delta^{\alpha\gamma}\delta^{\beta\delta}
 -\delta^{\alpha\delta}\delta^{\beta\gamma} \big)
 \big( \delta_{ac}\delta_{bd}-\delta_{ad}\delta_{bc}\big)\,
 \left(\psi^\alpha_a\sigma_2\psi^\beta_b\right)
 \left(\psi^{\gamma\,\dagger}_c\sigma_2 \psi^{\delta\,\dagger}_d\right).
\ee
The free energy of the system is the sum of the quasi-particle
contribution equ.~(\ref{evals}) and the mean field potential
$V=1/K\cdot(\Delta_{ud}^2+2\Delta_{us}^2)$. There are two coupled
gap equations, which can be derived by varying the free energy
with respect to the two parameters $\Delta_{ud}$ and $\Delta_{us}$.
A typical numerical result is shown in Fig.~\ref{fig_unlock}. 
We observe that the flavor symmetry breaking difference 
$\Delta_{ud}-\Delta_{us}$ is quite small all the way up to 
the critical strange quark mass. At the critical mass, there 
is a discontinuous transition to a phase where $\Delta_{us}$ 
vanishes exactly. The value of the critical mass is very close 
to the estimate $m_s=2\sqrt{\mu\Delta_{us}^0}$. 

 A number of authors have improved on the treatment presented 
in this section, in particular by studying the consequences
of imposing electric and color charge neutrality 
\cite{Steiner:2002gx,Alford:2002kj,Neumann:2002jm}. In the 
BCS framework one finds that once charge neutrality is imposed  
the number density of up, down and strange quarks in the CFL 
phase is exactly the same, even in the presence of flavor 
symmetry breaking. As a consequence, the CFL phase does not 
require the presence of electrons to be electrically neutral 
\cite{Rajagopal:2000ff}.

 Another interesting question concerns the possibility of 
additional phases that interpolate between the CFL and $N_f=2$ 
(2SC) superfluids. In \cite{Shovkovy:2003uu,Gubankova:2003uj} 
it was shown that the charge neutrality constraint may
help to stabilize gapless (`breached') CFL phases for 
$m_s>m_s({\it crit})$. Also, if the mismatch between 
the strange and non-strange Fermi surfaces is too large
for BCS pairing to occur, pairing may still take place
with a spatially varying superfluid order parameter
\cite{Alford:2000ze}. This phase is known as the LOFF
(Larkin-Ovchinikov-Fulde-Ferell) phase 
\cite{Larkin:1964,Fulde:1964,Abrikosov:1988}.
The LOFF phase is distinguished by an interesting 
crystal structure \cite{Bowers:2002xr}. 

 All of these studies are based on an ansatz for the 
structure of the CFL phase in the presence of flavor symmetry 
breaking. This is somewhat unsatisfactory. In weak coupling
and in the limit $m_s\ll m_s({\it crit})$ we should be able 
to perform rigorous calculations. Also, in the BCS approximation
we find that the quark densities in the CFL phase remain 
exactly equal even if the strange quark mass or the electron 
chemical potential are non-zero. However, the CFL phase contains 
almost massless flavored Goldstone bosons, so the response 
to any external perturbation that can couple to Goldstone 
modes should not vanish. 

 In the following section we will show how to study the
effect of a non-zero strange mass using an effective 
field theory of the CFL phase \cite{Casalbuoni:1999wu}. 
This theory determines both the ground state and the 
spectrum of excitations with energies below the gap in 
the CFL phase. Using the effective theory allows us to perform 
systematic calculations order by order in the quark mass. 

\begin{figure}[t]
\includegraphics[width=10cm]{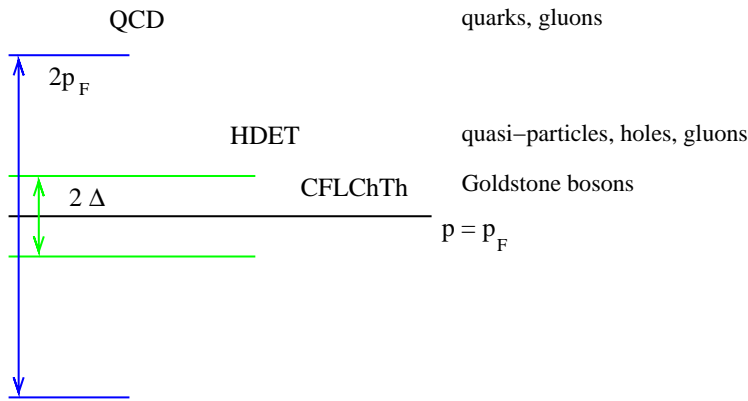}
\caption{\label{fig_eft}
Hierarchy of effective field theories in the CFL phase.}
\end{figure}

\subsection{CFL chiral theory}
\label{sec_CFLchi}

 For excitation energies smaller than the gap the only 
relevant degrees of freedom are the Goldstone modes 
associated with the breaking of chiral symmetry and
baryon number, see Fig.~\ref{fig_eft}. The interaction 
of the Goldstone modes is described by an effective 
lagrangian of the form \cite{Casalbuoni:1999wu}
\bea
\label{l_cheft}
{\cal L}_{eff} &=& \frac{f_\pi^2}{4} {\rm Tr}\left[
 \nabla_0\Sigma\nabla_0\Sigma^\dagger - v_\pi^2
 \partial_i\Sigma\partial_i\Sigma^\dagger \right] 
 +\Big[ B {\rm Tr}(M\Sigma^\dagger) + h.c. \Big] \\ 
 & & \hspace*{-1cm}\mbox{} 
     +\Big[ A_1{\rm Tr}(M\Sigma^\dagger)
                        {\rm Tr} (M\Sigma^\dagger) 
     + A_2{\rm Tr}(M\Sigma^\dagger M\Sigma^\dagger)   
     + A_3{\rm Tr}(M\Sigma^\dagger){\rm Tr} (M^\dagger\Sigma)
         + h.c. \Big]+\ldots . 
 \nonumber 
\eea
Here $\Sigma=\exp(i\phi^a\lambda^a/f_\pi)$ is the chiral field,
$f_\pi$ is the pion decay constant and $M$ is a complex mass
matrix. The chiral field and the mass matrix transform as
$\Sigma\to L\Sigma R^\dagger$ and  $M\to LMR^\dagger$ under 
chiral transformations $(L,R)\in SU(3)_L\times SU(3)_R$. We 
have suppressed the singlet fields associated with the breaking 
of the exact $U(1)_V$ and approximate $U(1)_A$ symmetries. As 
with ordinary chiral perturbation theory, the structure of the 
effective lagrangian is entirely determined by the symmetries.
At low density the coefficients $f_\pi$, $B,A_i,\ldots$ are 
non-perturbative quantities that have to extracted from 
experiment or measured on the lattice. At large density, on
the other hand, the chiral coefficients can be calculated in 
perturbative QCD. 

 Superficially, equ.~(\ref{l_cheft}) looks exactly like ordinary 
chiral perturbation theory. There are, however, some important 
differences. Lorentz invariance is broken and Goldstone modes 
move with the velocity $v_\pi<c$. The chiral expansion has 
the structure
\be
{\cal L}\sim f_\pi^2\Delta^2 \left(\frac{\partial_0}{\Delta}\right)^k
 \left(\frac{\vec{\partial}}{\Delta}\right)^l
  \big(\Sigma\big)^m\big(\Sigma^\dagger\big)^n.
\ee
Loop graphs are suppressed by powers of $p/(4\pi f_\pi)$.
We shall see that the pion decay constant scales as $f_\pi
\sim p_F$. As a result loops are suppressed by $p/p_F$
whereas higher order contact terms are suppressed by $p/\Delta$.
This means that in the CFL chiral theory pion loops with leading 
order vertices are parametrically small as compared to higher 
order contact terms, whereas in ordinary chiral perturbation
theory the two are comparable in size. 

  Further differences as compared to chiral perturbation theory 
in vacuum appear when the expansion in the quark mass is considered. 
The CFL phase has an approximate $(Z_2)_A$ symmetry under which 
$M\to -M$ and $\Sigma\to \Sigma$. This symmetry implies that the 
coefficients of mass terms that contain odd powers of $M$ are 
small. The $(Z_2)_A$ symmetry is explicitly broken by instantons.
The coefficient $B$ can be determined from a weak coupling 
instanton calculation and $B\sim (\Lambda_{QCD}/p_F)^{8}$, see
Sect.~\ref{sec_inst}. 

A priori it is also not clear what the expansion parameter in 
the chiral expansion is. There are several dimensionless 
ratios that can appear, $(m/p_F),(m/\Delta)$ and $(m/\Lambda_{QCD})$.
The BCS calculations discussed in the previous section suggest 
that the CFL phase undergoes a phase transition to a less 
symmetric phase when $m^2/(2p_F)\sim \Delta$. This result 
indicates that the expansion parameter is $M^2/(p_F\Delta)$. 
We shall see that this is indeed the case. However, the 
coefficients $A_{i}$ of the quadratic terms in $M$ turn 
out to be anomalously small. In Sect.~\ref{sec_hdet} we 
will show that
\be
 A_i M^2 \sim \Delta^2 M^2 \sim f_\pi^2\Delta^2 
\left(\frac{M^2}{p_F^2}\right), 
\ee
compared to the naive estimate $A_iM^2\sim f_\pi^2\Delta^2
[M^2/(p_F\Delta)]$.

 The pion decay constant $f_\pi$ and the coefficients $A_i$ can 
be determined using matching techniques. Matching expresses the 
requirement that Green functions in the effective chiral theory 
and the underlying microscopic theory, QCD, agree. The pion decay 
constant is most easily determined by coupling $SU(N_f)_{L,R}$
gauge fields $W_{L,R}$ to the left and right flavor currents. 
As usual, this amounts to replacing ordinary derivatives by 
covariant derivatives. The time component of the covariant 
derivative is given by $\nabla_0\Sigma=\partial_0 \Sigma+
iW_L\Sigma-i\Sigma W_R$ where we have suppressed the vector 
index of the gauge fields. In the CFL vacuum $\Sigma=1$ the 
axial gauge field $W_L-W_R$ acquires a mass by the Higgs 
mechanism. From equ.~(\ref{l_cheft}) we get
\be
\label{wm2}
{\cal L} = \frac{f_\pi^2}{4} \, \frac{1}{2} (W_L-W_R)^2.
\ee
The coefficients $B$ and $A_{1,2,3}$ can be determined by computing
the shift in the vacuum energy due to non-zero quark masses
in both the chiral theory and the microscopic theory. In the 
chiral theory we have 
\be 
\label{cfl_m}
\Delta{\cal E}=  
 -\Big[ B {\rm Tr}(M) + 
        A_1\left({\rm Tr}(M)\right)^2
      + A_2{\rm Tr}(M^2) + A_3{\rm Tr}(M){\rm Tr} (M^\dagger)
         + h.c. \Big].
\ee
We note that as long as we keep track of the difference
between $M$ and $M^\dagger$ different $O(M^2)$ mass terms 
produce distinct contributions to the vacuum energy. This 
means that the coefficients $A_i$ can be reconstructed 
uniquely from the vacuum energy. 

\begin{figure}[t]
\includegraphics[width=9cm]{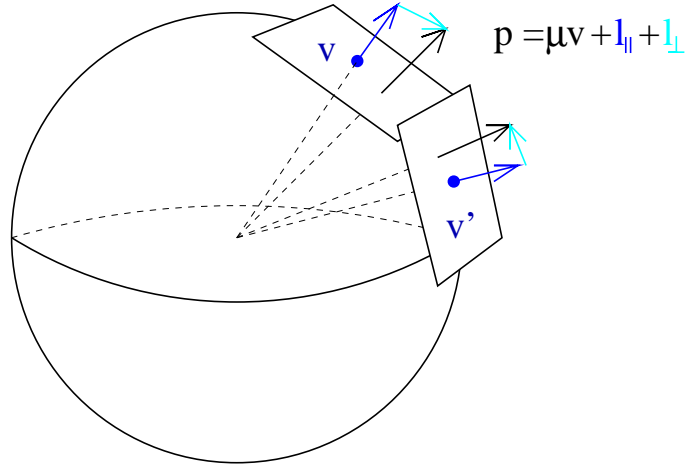}
\caption{\label{fig_hdet}
High density effective field theory description of the 
Fermi surface.}
\end{figure}

\subsection{High density effective theory}
\label{sec_hdet}

 In this section we shall determine the mass of the gauge field
and the shift in the vacuum energy in the CFL phase of QCD
at large baryon density. This is possible because asymptotic
freedom guarantees that the effective coupling is weak. The 
QCD Lagrangian in the presence of a chemical potential
is given by
\be
\label{qcd}
 {\cal L} = \bar\psi \left( i\Dslash +\mu\gamma_0 \right)\psi
 -\bar\psi_L M\psi_R - \bar\psi_R M^\dagger \psi_L 
 -\frac{1}{4}G^a_{\mu\nu}G^a_{\mu\nu},
\ee
where $D_\mu=\partial_\mu+igA_\mu$ is the covariant derivative,
$M$ is the mass matrix and $\mu$ is the baryon chemical 
potential. If the baryon density is very large perturbative QCD 
calculations can be further simplified. The main observation 
is that the relevant degrees of freedom are particle and hole 
excitations in the vicinity of the Fermi surface. We shall describe 
these excitations in terms of the field $\psi_+(\vec{v},x)$, where 
$\vec{v}$ is the Fermi velocity. The field $\psi_+(\vec{v},x)$
is defined on patches that cover the Fermi surface, see 
Fig.~\ref{fig_hdet}. Soft collinear scatterings take place 
within a given patch whereas hard interactions can scatter 
Fermions from one patch to another.

 At tree level, the quark field $\psi$ can be decomposed as 
$\psi=\psi_++\psi_-$ where $\psi_\pm=\frac{1}{2}(1\pm\vec{\alpha}
\cdot\hat{v})\psi$. To leading order in $1/p_F$ we can eliminate 
the field $\psi_-$ using its equation of motion. For $\psi_{-,L}$ 
we find
\be
\psi_{-,L} = \frac{1}{2p_F}
  \left( i\vec{\alpha}_\perp\cdot\vec{D}\psi_{+,L}
         + \gamma_0 M \psi_{+,R}\right).
\ee
There is a similar equation for $\psi_{-,R}$. The longitudinal 
and transverse components of $\gamma_\mu$ are defined by $(\gamma_0,
\vec{\gamma})_{\|}=(\gamma_0,\vec{v}(\vec{\gamma}\cdot\vec{v}))$ 
and $(\gamma_\mu)_\perp = \gamma_\mu-(\gamma_\mu)_{\|}$. To leading 
order in $1/p_F$ the lagrangian for the $\psi_+$ field is given by
\cite{Hong:2000tn,Hong:2000ru,Beane:2000ms}
\bea
\label{hdet}
{\cal L} &=& 
 \psi_{L+}^\dagger (iv\cdot D) \psi_{L+}
  - \frac{ \Delta}{2}\left(\psi_{L+}^{ai} C \psi_{L+}^{bj}
 \left(\delta_{ai}\delta_{bj}-
           \delta_{aj}\delta_{bi} \right) 
           + {\rm h.c.} \right) \nonumber \\ 
& & \hspace{0.5cm}\mbox{}
  - \frac{1}{2p_F} \psi_{L+}^\dagger \left(  (\Dslash_\perp)^2 
  + MM^\dagger \right)  \psi_{L+}  
  + \left( R\leftrightarrow L, M\leftrightarrow M^\dagger \right)  
  + \ldots ,
\eea
with $v_\mu=(1,\vec{v})$ and $i,j,\ldots$ and $a,b,\ldots$ denote 
flavor and color indices.  In order to perform perturbative 
calculations in the superconducting phase we have added a tree 
level gap term $\psi^{ai}_{L,R} C\Delta_{ai,bj} \psi^{bj}_{L,R}$. 
In the CFL phase this term has the structure $\Delta_{ai,bj}
=\Delta(\delta_{ai}\delta_{bj}-\delta_{aj}\delta_{bi})$. The 
magnitude of the gap $\Delta$ is determined order by order in
perturbation theory from the requirement that the thermodynamic
potential is stationary with respect to $\Delta$. With the gap 
term included the perturbative expansion is well defined.

\begin{figure}
\includegraphics[width=16cm]{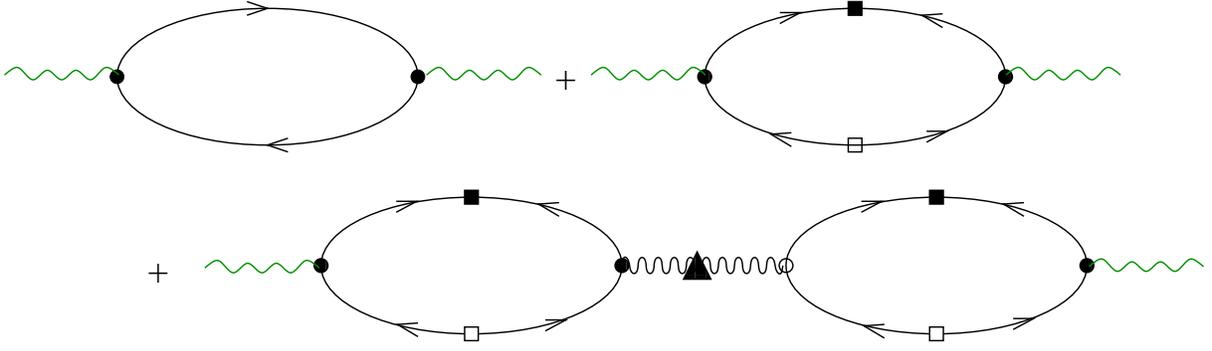}
\vspace*{0.5cm}
\caption{\label{fig_fpi}
Diagrams in the high density effective theory that contribute
to the chiral flavor current polarization function at leading 
order in $g$. The green wiggly lines are flavor currents, while
the black wiggly line is a gluon propagator. Squares are 
anomalous fermion self energy insertions, and the triangle
is a gluon self energy insertion.}
\end{figure}

  The screening mass of the flavor gauge fields $W_{L,R}$ 
can be determined by computing the corresponding polarization 
function in the limit $q_0=0$, $\vec{q}\to 0$. The relevant
diagrams are shown in Fig.~\ref{fig_fpi}, see 
\cite{Son:1999cm,Bedaque:2001je}. The first two diagrams 
do not involve mixing between left and right-handed 
currents. The third diagram involves mixing between 
left and right handed currents and is unique to the CFL
phase. We find
\be 
\label{pi_cfl}
\Pi^{LL}_{00}=\Pi^{RR}_{00}=-\Pi^{LR}_{00}=\frac{m_D^2}{4},
\hspace{0.5cm}
m_D^2=\frac{21-8\log(2)}{18}
 \left(\frac{p_F^2}{2\pi^2}\right).
\ee  
Matching equ.~(\ref{pi_cfl}) against equ.~(\ref{wm2}) we 
get \cite{Son:1999cm,Zarembo:2000pj,Miransky:2001bd} 
\be
\label{cfl_fpi}
f_\pi^2 = \frac{21-8\log(2)}{18} 
  \left(\frac{p_F^2}{2\pi^2} \right) .
\ee 
Repeating the matching calculation for the spatial components 
of the polarization tensor we get $v_\pi^2=1/3$ \cite{Son:1999cm}.

Our next task is to compute the mass dependence of the vacuum 
energy. To leading order in $1/p_F$ there is only one operator 
in the high density effective theory
\be 
\label{kin}
{\cal L} = -\frac{1}{2p_F} \left( \psi_{L+}^\dagger MM^\dagger \psi_{L+}
 + \psi_{R+}^\dagger M^\dagger M\psi_{R+} \right).
\ee
This term arises from expanding the kinetic energy of a massive
fermion around $p=p_F$. We note that $MM^\dagger/(2p_F)$ and
$M^\dagger M/(2p_F)$ act as effective chemical potentials for 
left and right-handed fermions, respectively. Indeed, to leading 
order in the $1/p_F$ expansion, the Lagrangian equ.~(\ref{hdet}) 
is invariant under a time dependent flavor symmetry $\psi_{L} 
\to L(t)\psi_{L}$, $\psi_{R}\to R(t)\psi_{R}$ where $X_L=
MM^\dagger/(2p_F)$ and $X_R=M^\dagger M/(2p_F)$ transform
as left and right-handed flavor gauge fields. If we impose 
this approximate gauge symmetry on the CFL chiral theory we 
have to include the effective chemical potentials 
$X_{L,R}$ in the covariant derivative of the chiral 
field \cite{Bedaque:2001je}, 
\be
\label{mueff}
 \nabla_0\Sigma = \partial_0 \Sigma 
 + i \left(\frac{M M^\dagger}{2p_F}\right)\Sigma
 - i \Sigma\left(\frac{ M^\dagger M}{2p_F}\right) .
\ee
$X_L$ and $X_R$ contribute to the vacuum energy at $O(M^4)$
\be
\label{E_m4}
\Delta {\cal E} = \frac{f_\pi^2}{8p_F^2} 
 {\rm Tr}\left[(MM^\dagger)(M^\dagger M)-(MM^\dagger)^2\right].
\ee
This result can also be derived directly in the microscopic 
theory \cite{Bedaque:2001je}. The corresponding diagrams are
exactly the same diagrams that appear in the calculation
of $f_\pi$, Fig.~\ref{fig_fpi}, but with the external flavor
gauge fields replaced by insertions of equ.~(\ref{kin}).
We also note that equation (\ref{E_m4}) has the expected 
scaling behavior ${\cal E}\sim f_\pi^2\Delta^2 [M^2/(p_F\Delta)]^2$.

\begin{figure}[t]
\includegraphics[width=16.5cm]{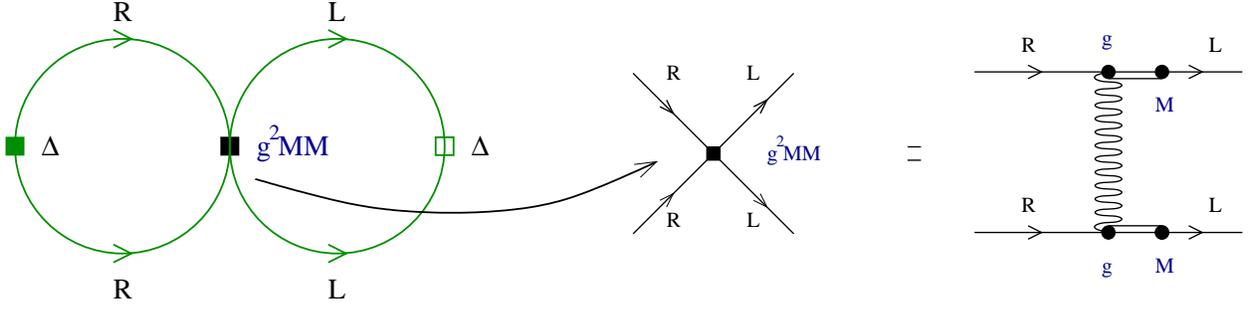}
\vspace*{0.5cm}
\caption{\label{fig_4f}
The left panel shows the diagram in the high density effective 
theory that contributes to the vacuum energy at order $O(M^2)$.
The right panel shows how the effective four-fermion vertex
can be matched against a chirality violating scattering 
amplitude involving a hard gluon exchange.}
\end{figure}

 $O(M^2)$ terms in the vacuum energy are generated by terms in 
the high density effective theory that are higher order in the 
$1/p_F$ expansion. These terms can be determined by computing 
chirality violating quark-quark scattering amplitudes for fermions 
in the vicinity of the Fermi surface \cite{Schafer:2001za}. Feynman 
diagrams for $q_L+q_L\to q_R+q_R$ are shown in Fig.~\ref{fig_4f}b. 
To leading order in the $1/p_F$ expansion the chirality violating 
scattering amplitudes are independent of the scattering angle and 
can be represented as local four-fermion operators
\be
\label{hdet_m}
 {\cal L} = \frac{g^2}{8p_F^4}
 \left( ({\psi^A_L}^\dagger C{\psi^B_L}^\dagger)
        (\psi^C_R C \psi^D_R) \Gamma^{ABCD} +
        ({\psi^A_L}^\dagger \psi^B_L) 
        ({\psi^C_R}^\dagger \psi^D_R) \tilde{\Gamma}^{ACBD} \right).
\ee
There are two additional terms with $(L\leftrightarrow R)$ and
$(M\leftrightarrow M^\dagger)$. We have introduced the CFL 
eigenstates $\psi^A$ defined by $\psi^a_i=\psi^A (\lambda^A)_{ai}
/\sqrt{2}$, $A=0,\ldots,8$. The tensor $\Gamma$ is defined by
\bea 
 \Gamma^{ABCD} &=& \frac{1}{8}\Big\{ {\rm Tr} \left[ 
    \lambda^A M(\lambda^D)^T \lambda^B M (\lambda^C)^T\right] 
  \nonumber \\
 & & \hspace{1cm}\mbox{}
   -\frac{1}{3} {\rm Tr} \left[
    \lambda^A M(\lambda^D)^T \right]
    {\rm Tr} \left[
    \lambda^B M (\lambda^C)^T\right] \Big\}.
\eea
The second tensor $\tilde{\Gamma}$ involves both $M$ and $M^\dagger$
and only contributes to terms of the form ${\rm Tr}[MM^\dagger]$ 
in the vacuum energy. These terms do not contain the chiral
field $\Sigma$ and therefore do not contribute to the masses
of Goldstone modes. We can now compute the shift in the vacuum 
energy due to the effective vertex equ.~(\ref{hdet_m}). The 
leading contribution comes from the two-loop diagram shown in 
Fig.~\ref{fig_4f}. This diagram is proportional to the square
of the superfluid density. We find
\be
\label{E_MM}
\Delta {\cal E} = -\frac{3\Delta^2}{4\pi^2} 
 \left\{  \Big( {\rm Tr}[M]\Big)^2 -{\rm Tr}\Big[ M^2\Big]
   \right\}
 + \Big(M\leftrightarrow M^\dagger \Big).
\ee
Note that a factor $g^2$ in the vertex equ.~(\ref{hdet_m}) 
cancels against a factor $1/g$ that comes from a logarithmic
$\log(\Delta)$ divergence in the superfluid density. Matching
equ.~(\ref{E_MM}) against equ.~(\ref{cfl_m}) we can determine the 
coefficients $A_{1,2,3}$. We find \cite{Son:1999cm,Schafer:2001za}
\be
 A_1= -A_2 = \frac{3\Delta^2}{4\pi^2}, 
\hspace{1cm} A_3 = 0.
\ee
We note that ${\cal E}\sim f_\pi^2\Delta^2 (\Delta/p_F)$
$[M^2/(p_F\Delta)]$ which shows that the coefficients $A_i$ 
are suppressed by $(\Delta/p_F)$. The effective lagrangian
equs.~(\ref{hdet}-\ref{hdet_m}) can also be used to compute 
higher order terms in $M$. The dominant $O(M^4)$ term is the
effective chemical potential term equ.~(\ref{E_m4}). Other
$O(M^4)$ terms are suppressed by additional powers of $(\Delta/
p_F)$.

\subsection{Instanton effects}
\label{sec_inst}

  In the CFL phase spontaneous chiral symmetry breaking
is dominated by order parameters of the form $\langle
(\bar{\psi}\psi)^2\rangle$ and as a consequence the 
coefficient of the linear mass term in the chiral lagrangian 
is suppressed as compared to the quadratic mass term.
Indeed, the CFL phase has an approximate $(Z_2)_A$ symmetry
that forbids the linear mass term. The $(Z_2)_A$ symmetry
is broken by instantons. This means that in the CFL
phase the quark-anti-quark condensate is induced by 
instantons.

\begin{figure}[t]
\includegraphics[width=12cm]{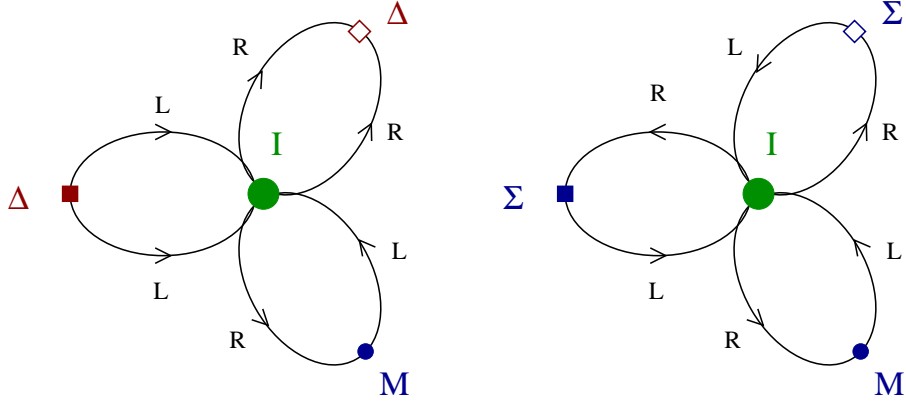}
\caption{\label{fig_ivac}
Instanton contribution to the $O(M)$ term in the vacuum energy.
The diagram on the left shows the instanton term in the CFL
phase at high baryon density. The squares are anomalous 
self energy insertions. The diagram on the right is the
corresponding term in QCD at low baryon density. The 
squares are normal self energy insertions.}
\end{figure}

 We shall determine the coefficient $B$ of the linear 
mass term by computing the instanton induced shift in 
the vacuum energy. The corresponding diagram is shown 
in Fig.~\ref{fig_ivac}. In QCD with three flavors, the 
instanton induced interaction between quarks is given 
by \cite{'tHooft:up,Shifman:uw,Schafer:1996wv}
\bea
\label{l_nf3}
{\cal L} &=& \int n(\rho,\mu)d\rho\, 
 \frac{(2\pi\rho)^6\rho^3}{6N_c(N_c^2-1)}
 \epsilon_{f_1f_2f_3}\epsilon_{g_1g_2g_3}
 \left( \frac{2N_c+1}{2N_c+4}
  (\bar\psi_{R,f_1} \psi_{L,g_1})
  (\bar\psi_{R,f_2} \psi_{L,g_2})
  (\bar\psi_{R,f_3} \psi_{L,g_3}) \right. \nonumber \\
& & \mbox{}\left. - \frac{3}{8(N_c+2)}
  (\bar\psi_{R,f_1} \psi_{L,g_1})
  (\bar\psi_{R,f_2} \sigma_{\mu\nu} \psi_{L,g_2})
  (\bar\psi_{R,f_3} \sigma_{\mu\nu} \psi_{L,g_3})
  + ( L \leftrightarrow R ) \right)  .
\eea
Here, $\rho$ is the instanton size, $\mu$ is the quark chemical
potential, $f_i,g_i$ are flavor indices and $\sigma_{\mu\nu}
=\frac{i}{2}[\gamma_\mu,\gamma_\nu]$. The instanton size
distribution $n(\rho,\mu)$ is given by
\bea
\label{G_I}
  n(\rho,\mu) &=& C_{N} \ \left(\frac{8\pi^2}{g^2}\right)^{2N_c} 
 \rho^{-5}\exp\left[-\frac{8\pi^2}{g(\rho)^2}\right]
 \exp\left[-N_f\rho^2\mu^2\right],\\
 && C_{N} \;=\; \frac{0.466\exp(-1.679N_c)1.34^{N_f}}
    {(N_c-1)!(N_c-2)!}\, ,\\
 && \frac{8\pi^2}{g^2(\rho)} \;=\; 
    -b\log(\rho\Lambda), \hspace{1cm} 
    b = \frac{11}{3}N_c-\frac{2}{3}N_f \, . 
\eea
At zero density, the $\rho$ integral in equ.~(\ref{G_I}) is
divergent at large $\rho$. This is the well-known infrared 
problem of the semi-classical approximation in QCD. At 
large chemical potential, however, large instantons 
are suppressed and the typical instanton size is $\rho
\sim \mu^{-1} \ll \Lambda^{-1}$. We also note that at
zero density the effective lagrangian equ.~(\ref{l_nf3})
is derived by computing $U(1)_A$ violating Greens functions
in the limit $p\ll\rho^{-1}$. From a similar study at finite 
density one can show that at $\mu\neq0$ equ.~(\ref{l_nf3})
has to be interpreted as an effective lagrangian for 
momenta in the vicinity of the Fermi surface $|\vec{p}|
-p_F\ll \rho^{-1} \sim \mu$ 
\cite{Schafer:1998up,Carter:1999mt,Schafer:2002ty}.

 To linear order in the quark mass one of the three zero 
modes is lifted. We find
\bea
\label{l_nf2}
{\cal L} &=& \int n(\rho,\mu)d\rho\, 
   \frac{2(2\pi\rho)^4\rho^3}{4(N_c^2-1)}
 \epsilon_{f_1f_2f_3}\epsilon_{g_1g_2g_3}M_{f_3 g_3}
 \left( \frac{2N_c-1}{2N_c}
  (\bar\psi_{R,f_1} \psi_{L,g_1})
  (\bar\psi_{R,f_2} \psi_{L,g_2}) \right. \\
& & \hspace{1cm}\mbox{}\left. - \frac{1}{8N_c}
  (\bar\psi_{R,f_1} \sigma_{\mu\nu} \psi_{L,g_1})
  (\bar\psi_{R,f_2} \sigma_{\mu\nu} \psi_{L,g_2})
  + (M\leftrightarrow M^\dagger, 
     L \leftrightarrow R ) \right) \nonumber ,
\eea
We can now compute the expectation value of equ.~(\ref{l_nf2}) in 
the CFL ground state \cite{Schafer:1999fe,Rapp:1999qa,Son:2001jm}. 
To leading order in perturbative QCD we can use the mean field
approximation $\langle (\bar{\psi}\psi)(\bar{\psi}\psi)\rangle
\sim \langle \psi\psi\rangle \langle\bar\psi\bar\psi\rangle$.
The instanton contribution to the vacuum energy density is
\bea
\label{E_I}
{\cal E} &=& -\int n(\rho,\mu) d\rho\,
 \frac{16}{3}(\pi\rho)^4\rho^3 
 \left[\frac{3\sqrt{2}\pi}{g}\Delta
     \left(\frac{\mu^2}{2\pi^2}\right)\right]^2
  {\rm Tr}\left[M+M^\dagger\right] ,
\eea
where we have used the perturbative result for the diquark 
condensate in the CFL phase
\bea
\label{qq_cond}
\langle \psi^a_{L,f} C\psi^b_{L,g}\rangle 
  &=& -\langle \psi^a_{R,f} C\psi^b_{R,g}\rangle = 
 \left( \delta^a_f\delta^b_g-\delta^a_g\delta^b_f \right) \Phi,\\
 & & \Phi \,=\,   \frac{3\sqrt{2}\pi}{g} \Delta 
     \left(\frac{\mu^2}{2\pi^2}\right). \nonumber
\eea 
We note that for $M={\rm diag}(m_u,m_d,m_s)$ the instanton
contribution to the vacuum energy is indeed negative. Since the
effective interaction involves both left and right-handed 
fermions the relative phase between the left and right-handed
condensate in equ.~(\ref{qq_cond}) is important. Instantons 
favor the state with $\langle\psi_L\psi_L\rangle = -\langle
\psi_R\psi_R\rangle$ which is the parity even ground state. 
Equation (\ref{E_I}) for the vacuum energy can be matched against 
the effective lagrangian equ.~(\ref{cfl_m}). We find
\be 
\label{B}
 B = C_N\frac{8\pi^4}{3}\frac{\Gamma(6)}{3^6}
 \left[\frac{3\sqrt{2}\pi}{g}\Delta
     \left(\frac{\mu^2}{2\pi^2}\right)\right]^2
  \left(\frac{8\pi^2}{g^2}\right)^{6}
  \left(\frac{\Lambda}{\mu}\right)^{12}\Lambda^{-3},
\ee
where we have performed the integral over the instanton
size $\rho$ using the one-loop beta function. The coefficient
$B$ is related to the quark-anti-quark condensate, $\langle
\bar{\psi}\psi\rangle =-2B$. We note that $B$ is indeed
parametrically small, $B\sim (\Lambda/p_F)^8$. 

\begin{figure}[t]
\includegraphics[width=9cm]{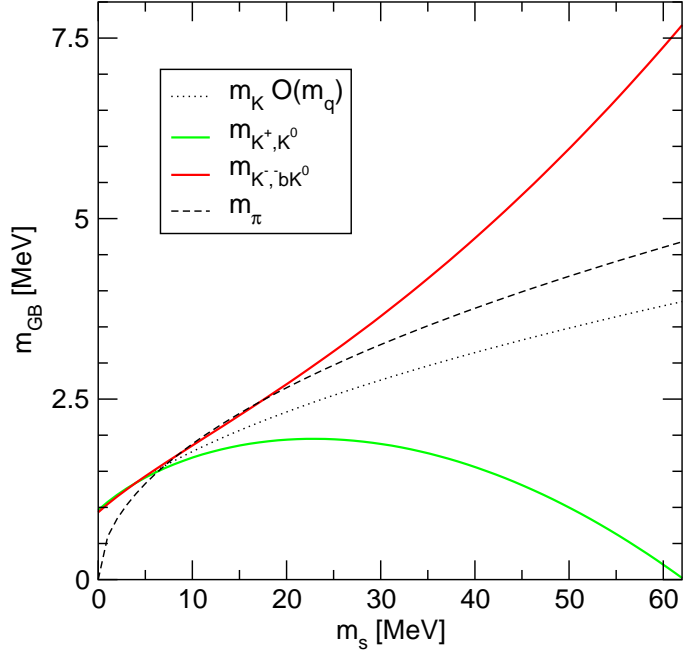}
\caption{\label{fig_kmass}
Goldstone boson masses in the CFL phase as a function
of the strange quark mass, from \cite{Bedaque:2001je}.}
\end{figure}

 In QCD with three flavors chiral symmetry is broken 
both at small and at large density. The instantons 
contribution to the linear Goldstone boson mass term
in the low density phase is shown in Fig.~\ref{fig_ivac}b.  
Again, the coefficient of the ${\rm Tr}(M\Sigma)$ term 
in the effective lagrangian is the instanton contribution 
to the quark condensate. There are strong arguments that 
instantons dominate chiral symmetry breaking in QCD at 
zero density, see \cite{Schafer:1996wv} for a review. 
This raises the question whether the instanton mechanisms 
of Fig.~\ref{fig_ivac}a and b are continuously connected. 
In a simple mean field calculation it was found that there 
is a first order phase transition that separates instanton 
induced chiral symmetry breaking at low and high density
\cite{Rapp:1999qa}, but this question certainly deserves 
further study.

\subsection{Kaon condensation}
\label{sec_kcond}

 Using the results discussed in the previous sections we
can compute the masses of Goldstone bosons in the CFL phase.
In Sect.~\ref{sec_CFLchi} we argued that the expansion parameter 
in the chiral expansion of the Goldstone boson masses is $\delta=
m^2/(p_F\Delta)$. The first term in this expansion comes from the 
$O(M^2)$ term in equ.~(\ref{l_cheft}), but the coefficients $A$ contain 
the additional small parameter $\epsilon=(\Delta/p_F)$. In a combined 
expansion in $\delta$ and $\epsilon$ the $O(\epsilon\delta)$ mass 
term and the $O(\delta^2)$ chemical potential term equ.~(\ref{mueff})
appear at the same order. Instanton effects are suppressed by 
extra powers of $(\Lambda/p_F)$. To order $O(\epsilon\delta,
\delta^2)$ the masses of the flavored Goldstone bosons are
\bea 
\label{mgb}
 m_{\pi^\pm} &=&  \mp\frac{m_d^2-m_u^2}{2p_F} +
         \left[\frac{4A}{f_\pi^2}(m_u+m_d)m_s\right]^{1/2},\nonumber \\
 m_{K_\pm}   &=&  \mp \frac{m_s^2-m_u^2}{2p_F} + 
         \left[\frac{4A}{f_\pi^2}m_d (m_u+m_s)\right]^{1/2}, \\
 m_{K^0,\bar{K}^0} &=&  \mp \frac{m_s^2-m_d^2}{2p_F} + 
         \left[\frac{4A}{f_\pi^2}m_u (m_d+m_s)\right]^{1/2}.\nonumber
\eea
Further studies of Goldstone boson properties in the CFL 
phase can be found in 
\cite{Rho:2000xf,Hong:2000ei,Manuel:2000wm,Rho:2000ww,Beane:2000ms}.
We observe that the pion masses are not strongly affected 
by the effective chemical potential $\mu_s=m_s^2/(2p_F)$
but the masses of the $K^+$ and $K^0$ are substantially lowered 
while the $K^-$ and $\bar{K}^0$ are pushed up. As a result the 
$K^+$ and $K^0$ meson become massless if 
\be
\label{ms_crit}
\left. m_s \right|_{crit}= 3.03\cdot  m_d^{1/3}\Delta^{2/3},
\ee
see Fig.~\ref{fig_kmass}. For larger values of $m_s$ the 
kaon modes are unstable, signaling the formation of a kaon 
condensate. 

\begin{figure}[t]
\includegraphics[width=12cm]{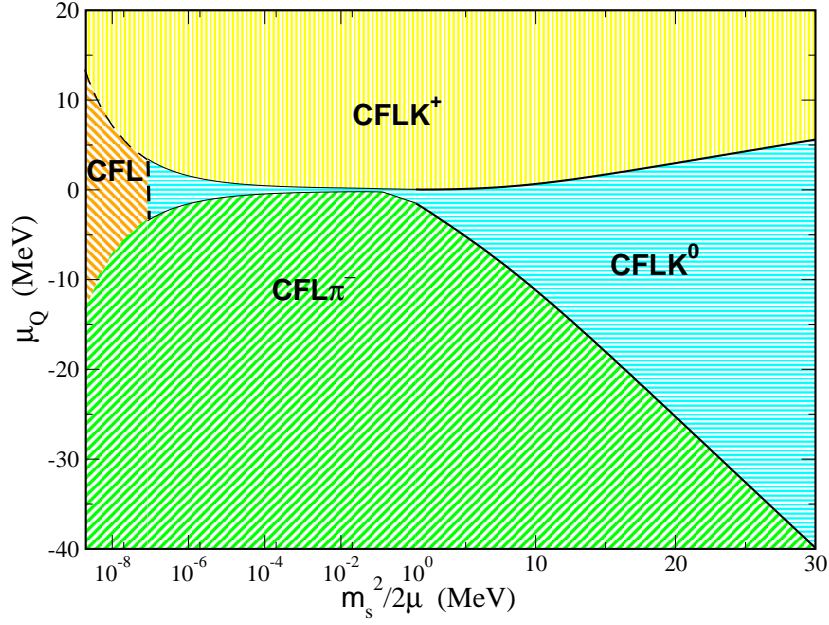}
\vspace*{0.5cm}
\caption{\label{fig_kcond}
This figure shows the phase structure of CFL matter as 
a function of the strange quark mass $m_s$ and the lepton
chemical potential $\mu_Q$, from \cite{Kaplan:2001qk}.}
\end{figure}

 Kaon condensation implies that the CFL ground state is reorganized.
For simplicity, we consider the case of exact isospin symmetry
$m_u=m_d\equiv m$. Kaon condensation can be studied using an 
ansatz of the form $\Sigma = \exp(i\alpha\lambda_4)$. The 
vacuum energy is 
\be 
\label{k0+_V}
 V(\alpha) = -f_\pi^2 \left( \frac{1}{2}\left(\frac{m_s^2-m^2}{2p_F}
   \right)^2\sin(\alpha)^2 + (m_{K}^0)^2(\cos(\alpha)-1)
   \right),
\ee
where $(m_K^0)^2= (4A/f_\pi^2)m_{u,d} (m_{u,d}+m_s)$ is the $O(M^2)$ 
kaon mass in the limit of exact isospin symmetry. Minimizing the vacuum 
energy we obtain $\alpha=0$ if $m_s^2/(2p_F)<m_K^0$ and $\cos(\alpha)
=(m_K^0)^2/\mu_s^2$ with $\mu_s=m_s^2/(2p_F)$ if $\mu_s >m_K^0$. 
In the kaon condensed phase $SU(2)_I\times U(1)_Y$ is spontaneously
broken to $U(1)_Q$. This coincides with the symmetry breaking 
pattern in the electroweak sector of the standard model. Kaon
condensation also provides an interesting realization of 
Goldstone's theorem. Even though the number of broken generators
is three, the number of Goldstone bosons is only two. This 
is related to the fact that one of the Goldstone modes has a 
quadratic dispersion relation \cite{Miransky:2001tw,Schafer:2001bq}.

 The hypercharge density in the kaon condensed phase is given by
\be 
\label{n_y}
n_Y = f_\pi^2 \mu_s \left( 1 -\frac{(m_K^0)^4}{\mu_s^4}\right).
\ee
This result is exactly analogous to the behavior of the charge
density in the case of the dilute Bose gas, equ.~(\ref{rho_bose}), 
and $N_c=2$ QCD, equ.~(\ref{rho_nc2}). We observe that within the 
range of validity of the effective theory, $\mu_s<\Delta$, the 
hypercharge density satisfies $n_Y<\Delta p_F^2/(2\pi^2)$. This 
means that the number of condensed kaons is bounded by the number 
of particles contained within a strip of width $\Delta$ around 
the Fermi surface. The maximum hypercharge density allowed by
the effective theory corresponds to the case that essentially
all strange quarks have been removed from the CFL wave function. 
This raises the question whether the CFL-2SC transition might 
be more complicated than what is suggested by the simple BCS 
model discussed in Sect.~\ref{sec_cfl_ms}, see Fig.~\ref{fig_unlock}.

 In the limit of exact isospin symmetry the $K^0$ and $K^+$
condensed phases are degenerate. If charge neutrality is 
imposed then the $K^0$ condensed phase is preferred. The 
effective lagrangian equ.~(\ref{l_cheft}) can be used to 
study a number of question related to the structure and 
the low energy properties of the CFL phase. The phase 
structure as a function of the strange quark mass and 
non-zero lepton chemical potentials was studied by Kaplan
and Reddy \cite{Kaplan:2001qk}, see Fig.~\ref{fig_kcond}. 
Other investigations have focused on the role of defects 
such as $K^0$ vortices \cite{Kaplan:2001hh,Buckley:2002ur},
as well as transport properties such as neutrino emissivity 
and thermal conductivity 
\cite{Reddy:2002xc,Jaikumar:2002vg,Shovkovy:2002kv}.

\section{Conclusion: The many phases of QCD}
\label{sec_sum}

\begin{figure}
\includegraphics[width=11.0cm]{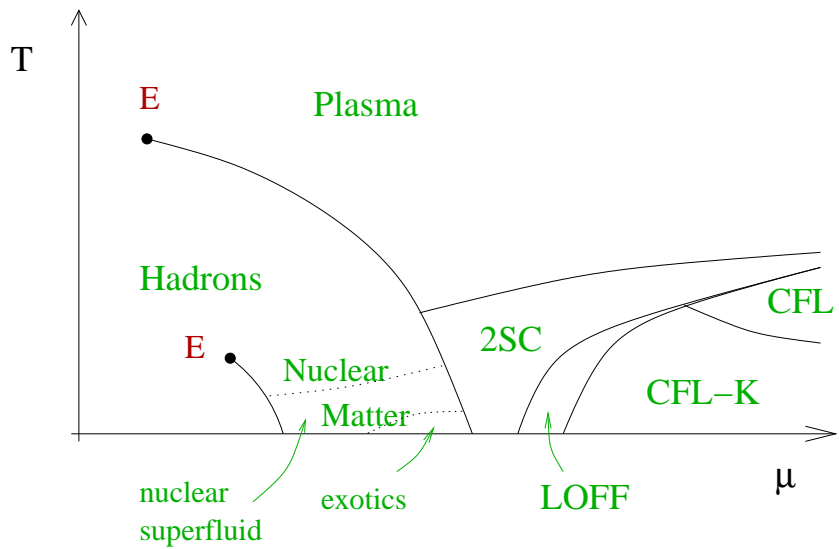}
\caption{\label{fig_phase_4}
Conjectured phase diagram of three flavor QCD with 
realistic quark masses.}
\end{figure}

 There are many issues in QCD at finite density that 
remain to be addressed. First of all it would clearly be 
of great help if some of the ideas discussed in these 
lectures could be tested using numerical simulations 
on the lattice. While some progress in this direction 
has been made, the most interesting regime of the phase 
diagram remains unexplored. Even more importantly, we
have to find experimental or observational constraints
on the properties of quark matter. In Sect.~\ref{sec_dqcd}
we mentioned the possibility of observing the QCD tri-critical
in relativistic heavy ion collisions. Observational
constraints on the properties of dense cold matter can
be obtained from neutron stars. We refer the reader 
to \cite{Reddy:2002ri} for a recent review and a guide
to the literature.

 On the more theoretical side we would like to improve 
our understanding of the phase diagram of QCD for 
realistic values of the strange quark mass. In 
Figs.~\ref{fig_phase_1},\ref{fig_phase_2},\ref{fig_phase_3}
we have shown proposed phase diagrams for idealized 
versions of QCD. A conjectured phase diagram for QCD
with three flavors and realistic quark masses is shown
in Fig.~\ref{fig_phase_4}. The reader will recognize 
many of the phases that we have discussed in the 
previous sections. However, from our discussion it is
also clear that many of the phases, the transition lines,
and the order of the transition are just guesses, based 
on mean-field or weak coupling arguments. Eventually, this 
problem will have to be resolved using observations and 
numerical simulations. In the mean time, however, we would 
also like to improve and systematize the theoretical 
approaches. This should be possible in both the limit 
of high and low density, using, respectively, weak coupling 
QCD and nuclear effective field theory.
  
 Acknowledgments: I would like to thank the 
organizers of the BARC workshop on ``Mesons and
Quarks'' for their hospitality. I would also like
to acknowledge the many discussions on dense QCD
I have had with friends and collaborators over the
years. I would like to thank, in particular, Mark
Alford, Paulo Bedaque, Krishna Rajagopal, Francesco
Sannino, Dam Son, Edward Shuryak, Misha Stephanov, and  
Frank Wilczek. This work was supported in part by a US DOE 
OJI grant. The write-up was completed during the INT workshop
on ``The first three years of running at RHIC''.
I would like to thank the INT in Seattle for hospitality.


\end{document}